\documentclass[letterpaper]{article} 
\usepackage{aaai2026}  
\usepackage{times}  
\usepackage{helvet}  
\usepackage{courier}  
\usepackage[hyphens]{url}  
\usepackage{graphicx} 
\usepackage{natbib}  
\usepackage{caption} 
\usepackage{algorithm}
\usepackage{algorithmic}

\usepackage{newfloat}
\usepackage{listings}
\DeclareCaptionStyle{ruled}{labelfont=normalfont,labelsep=colon,strut=off} 
\floatstyle{ruled}
\newfloat{listing}{tb}{lst}{}
\floatname{listing}{Listing}
\usepackage{booktabs}
\usepackage{array}
\usepackage{multirow}
\usepackage{rotating}
\usepackage{tikz}
\usetikzlibrary{positioning, shadows, shapes.geometric, fit, backgrounds, calc}
\usepackage[table]{xcolor}
\usepackage[switch]{lineno}
\nolinenumbers

\usepackage{xspace}
\usepackage{amsmath}
\allowdisplaybreaks
\usepackage{amssymb}
\usepackage{amsthm}

\newcommand{\buchi}{B\"uchi\xspace}

\newcommand{\lang}[1]{\mathcal{L}(#1)}

\usepackage{amsmath,nccmath}
\usepackage{amssymb}
\usepackage{xspace}
\usepackage{xcolor}
\usepackage{lineno}  

\usepackage{cancel} 

\usepackage[capitalise]{cleveref}

\newcommand{\af}{\mathit{af}} 

\newcommand{\Dist}{\mathrm{Distr}} 
\newcommand{\suchthat}{\;\ifnum\currentgrouptype=16 \middle\fi|\;} 

\newcommand{\A}{\mathcal{A}}

\newcommand{\D}{\mathcal{D}}
\newcommand{\E}{\mathcal{E}}

\newcommand{\pa}{\mathcal{P}}

\newcommand{\M}{\mathcal{M}} 
\newcommand{\N}{\mathcal{N}}

\newcommand{\R}{\mathcal{R}}
 
\newcommand{\T}{\mathcal{T}}

\newcommand{\NBA}{{\sc nba}\xspace}

\newcommand{\DBA}{{\sc dba}\xspace}

\newcommand{\ttrue}{{\mathit{tt}}}
\newcommand{\ffalse}{\mathit{ff}}

\newcommand{\ls}{{\sf{lst}}}

\newcommand{\alphabet}{\Sigma}

\newcommand{\finwords}{\alphabet^*}

\newcommand{\infwords}{\alphabet^\omega}

\newcommand{\states}{Q}

\newcommand{\trans}{\delta}

\newcommand{\init}{q_0}

\newcommand{\run}{\rho}
\newcommand{\langsymb}[0]{\mathcal{L}}
\newcommand{\finlang}[1]{\langsymb_{*}(#1)}
\newcommand{\inflang}[1]{\langsymb(#1)}
\newcommand{\flang}[1]{[#1]}
\newcommand{\size}[1]{|#1|}

\newcommand{\fpaths}{\rm{FPaths}}
\newcommand{\ipaths}{\rm{IPaths}}
\newcommand{\distr}{\mathsf{Distr}}

\newcommand{\ltl}{\text{LTL}\xspace}

\newcommand{\ltlfU}{\textsf{U}}
\newcommand{\ltlfX}{\textsf{X}}

\newcommand{\ltlfG}{\textsf{G}}

\newcommand{\ltlfF}{\textsf{F}}

\newcommand{\acccond}{\alpha}

\newcommand{\psem}{\textsf{Psem}}
\newcommand{\psyn}{\textsf{Psyn}}
\newcommand{\pp}{\mathbb{P}}

\newtheorem{theorem}{Theorem}
\newtheorem{lemma}{Lemma}

\newcommand{\ap}{\textsf{AP}}
\newcommand{\setnocond}[1]{\{#1\}}

\newcommand{\prob}{\mathsf{P}}
\newcommand{\act}{\text{Act}}
\newcommand{\lab}{\textit{L}}

\newcommand{\order}{\textsf{idx}}

\newcommand{\Agfm}{\mathcal{A}_\text{GFM}}

\newcommand{\Adba}{\mathcal{A}_\text{DBA}}

\newcommand{\Adca}{\mathcal{A}_\text{DCA}}

\newcommand{\Agfgmin}{\mathcal{A}_\text{GFG-Min}}

\newcommand{\Anba}{\mathcal{A}_\text{NBA}}

\newcommand{\Apa}{\mathcal{A}_\text{PA}}

\newcommand{\action}{a}
\newcommand{\strategy}{\mu}

\newcommand{\letter}{\sigma}

\renewcommand{\NBA}{\text{NBA}}

\usepackage[framemethod=TikZ]{mdframed}
\usepackage{tikz}
\usetikzlibrary{automata, positioning, arrows.meta}
\tikzset{
  >=Latex,
  acc/.style={
    preaction={draw, line width=2.6pt, draw=cyan!70!black}, 
    preaction={draw, line width=1.6pt, draw=white},         
    line width=0.9pt                                        
  }
}

\title{Good-for-MDP State Reduction for Stochastic LTL Planning}
\author {
    Christoph Weinhuber\textsuperscript{\rm 1},
    Giuseppe De Giacomo\textsuperscript{\rm 1},
    Yong Li\textsuperscript{\rm 2}\footnote{Corresponding author},
    Sven Schewe\textsuperscript{\rm 3},
    Qiyi Tang\textsuperscript{\rm 3}
}
\affiliations {
    \textsuperscript{\rm 1}University of Oxford, Oxford, UK\\
    \textsuperscript{\rm 2}Key Laboratory of System Software (Chinese Academy of Sciences), \\Institute of Software Chinese Academy of Sciences, PRC\\
    \textsuperscript{\rm 3}University of Liverpool, UK\\
    \{christoph.weinhuber, giuseppe.degiacomo\}@cs.ox.ac.uk, liyong@ios.ac.cn, \{sven.schewe, qiyi.tang\}@liverpool.ac.uk
}

\begin{document}\sloppy
\maketitle

\begin{abstract}
We study stochastic planning problems in Markov Decision Processes (MDPs) with goals specified in Linear Temporal Logic (LTL). The state-of-the-art approach transforms LTL formulas into good-for-MDP (GFM) automata, which feature a restricted form of nondeterminism. These automata are then composed with the MDP, allowing the agent to resolve the nondeterminism during policy synthesis.
A major factor affecting the scalability of this approach is the size of the generated automata. In this paper, we propose a novel GFM state-space reduction technique that significantly reduces the number of automata states. Our method employs a sophisticated chain of transformations, leveraging recent advances in good-for-games minimisation developed for adversarial settings.
In addition to our theoretical contributions, we present empirical results demonstrating the practical effectiveness of our state-reduction technique.
Furthermore, we introduce a direct construction method for formulas of the form $\ltlfG\ltlfF\varphi$, where $\varphi$ is a co-safety formula. This construction is provably single-exponential in the worst case, in contrast to the general doubly-exponential complexity. Our experiments confirm the scalability advantages of this specialised construction.

\end{abstract}

\section{Introduction}

Planning with temporal objectives has a long tradition in artificial intelligence. 
Early pioneering work, such as \cite{DBLP:conf/aaai/BacchusBG96, DBLP:conf/aaai/BacchusBG97, DBLP:conf/aaai/Littman97, DBLP:journals/jair/LittmanGM98, DBLP:journals/jair/ThiebauxGSPK06}, defined key formalisms like \emph{Markov decision processes} (MDPs), analysed the computational complexity of planning problems and introduced methods for handling \emph{non-Markovian rewards}.

These foundations paved the way for more expressive and structured approaches, most notably the use of \emph{Linear Temporal Logic} (\ltl)~\cite{pnueli1977temporal}, to formally specify complex goals \cite{DBLP:conf/iros/LacerdaPH14, DBLP:journals/ai/BrafmanG24}.
This field is now even pushing into areas like reinforcement learning \cite{DBLP:conf/ijcai/YangLC22}, multi-agent systems \cite{schillinger2019hierarchical}, and handling uncertainty \cite{DBLP:journals/corr/abs-2502-19603}.

Traditionally, in order to solve an MDP with an \ltl goal $\varphi$, we first build a \emph{nondeterministic \buchi automaton} (NBA) for $\varphi$.
Then we convert the NBA to an equivalent \emph{deterministic automaton}, e.g., a deterministic Rabin automaton (DRA)~\cite{DBLP:conf/focs/Safra88}.
Recent progresses in this direction include direct translations of LTL to DRAs~\cite{DBLP:conf/cav/EsparzaK14}.
Afterwards, we need to perform the Cartesian product of the MDP and the DRA, identify \emph{accepting end-components} (ECs) and compute a \emph{(memoryless) strategy} with maximal probability of reaching accepting ECs.

State-of-the-art approaches \cite{DBLP:conf/concur/HahnLST015,DBLP:conf/cav/SickertEJK16} proposed the usage of \emph{limit-deterministic \buchi automata} (LDBAs) to avoid the notorious problem of obtaining DRAs. 
LDBAs give the burden of resolving the \emph{nondeterminism} in automata to the agent, without changing the optimal satisfaction probability. 
Yet, only a certain type of LDBAs works for all finite MDPs. 
The precise criterion for NBAs is expressed by being \emph{``good-for-MDP"} (GFM), where all nondeterministic choices are \emph{angelic} in the sense that they can be resolved by the agent, while preserving the optimal satisfaction probability \cite{DBLP:conf/tacas/HahnPSS0W20}. While the LTL to GFM construction does not change the complexity of the overall planning problem, in practice, the difference among using GFMs and DRAs is significant \cite{DBLP:conf/cav/MeyerSL18}.

A major factor affecting the scalability of solving MDPs with \ltl goals is the size of the
GFM automata corresponding to the \ltl goals~\cite{DBLP:conf/concur/HahnLST015, DBLP:conf/atva/SickertK16}. In this paper, we propose a novel GFM
state-space reduction technique that significantly reduces the
number of automata states. 
Our method employs a sophisticated chain of transformations, which allows us to leverage recent advances in
\emph{good-for-games} minimisation developed for adversarial settings \cite{AbuRadiK22} to our context. 
We provide experimental evidence that our state-reduction technique effectively reduces the automata state space, by taking benchmark examples of \ltl goals from over eight sources, including influential papers in planning, verification and reinforcement learning.

Furthermore, we introduce a \emph{direct} method for constructing the GFM automata of formulas of the form $\ltlfG \ltlfF \varphi$, where $\varphi$ is a \emph{co-safety} property.
This kind of formulas is common in both verification \cite{holevcek2004verification} and reinforcement learning \cite{deepltl}. 
We show that the GFM automaton obtained through our specialised construction incurs only a \emph{singly} exponential blow-up, in contrast to the doubly exponential blow-up in the general case.
Moreover, we provide experimental evidence that this theoretical advantage indeed translates into significantly fewer states.

\section{Preliminaries}

\paragraph{Markov Decision Processes.}
Following \cite{DBLP:books/daglib/0020348},
a Markov decision process (MDP) $\M$ is a tuple $(S, \act, \alphabet, \prob, s_0, \lab)$ with a finite set of states $S$, a set of actions $\act$, a set of labels $\alphabet$, a transition
probability function $\prob : S \times \act \times S \rightarrow [0, 1]$, an initial state $s_0 \in S$ and a labelling function $\lab : S \times  \act \rightarrow \alphabet$ that labels state-action pairs to the set of propositions that hold in that state\footnote{This generalises the usual labelling function $\lab: S \rightarrow \alphabet$.}.
We abuse the notation by writing $\act(s)$ to denote the set of available actions at state $s$.
A path $\xi$ of $\M$ is an (in)finite sequence of alternating states and actions $\xi=s_0 \action_0 s_1 \action_1\cdots$, ending with a state if finite, such that for all $i 
\geq 0$, $\action_i \in \act(s_i)$ and $\prob(s_i, \action_i, s_{i+1}) > 0 $. 
The sequence $\lab(\xi)=\lab(s_0, \action_0)\lab(s_1,\action_1), \cdots$ over $\alphabet$ is called the \emph{trace} induced by the path $\xi$ over $\M$.
We denote by $\fpaths$ and $\ipaths$ the set of all finite and infinite paths of $\M$. 

A (finite-memory) strategy $\strategy$ of $\M$ is a function $\strategy: \fpaths \to \distr(\act)$ such that, for each $\xi\in \fpaths$, $\strategy(\xi)\in \distr(\act({\ls}(\xi)))$, where ${\ls}(\xi)$ is the last state of the finite path $\xi$ and $\distr(\act)$ denotes the set of all possible distributions over $\act$.
Let $\Omega_\strategy^{\M}(s_0)$ denote the subset of (in)finite paths of $\M$ that correspond to strategy $\strategy$ and initial state $s_0$.
$\strategy$ is \emph{memoryless} if $\strategy: S \rightarrow \Dist(\act)$, meaning it selects actions independently of the past.

A strategy $\strategy$ of $\M$ is able to resolve the nondeterminism of an MDP and induces a Markov chain (MC)
$\M^{\strategy} = (\fpaths, \alphabet, \prob_{\strategy},  \lab')$ where, for $\xi = s_0 \action_0 \cdots s_{n-1} \action_{n-1} s_n \in \fpaths$ and $\action_n \in \act(s_n)$, $\prob_{\strategy}(\xi, \xi\cdot \action_n \cdot s_{n+1}) = \prob(s_{n},\action_{n}, s_{n+1}) \cdot \strategy(\xi)(\action_{n})$ and $\lab'(\xi) = \lab(s_{n-1}, \action_{n-1})$.

A \emph{sub-MDP} of $\M$ is an MDP $\M' = (S', \act', \alphabet, \prob', \lab)$ where $S' \subseteq S, \act' \subseteq \act$ is such that, for every $s \in S'$, $\act'(s)\subseteq \act(s)$, and $\prob'$ and $\lab'$ are analogous to $\prob$ and $\lab$ when restricted to $S'$ and $\act'$.
In particular, $\M'$ is closed under probabilistic transitions, i.e., for all $s \in S'$ and $\action \in\act' $ we have that $\prob(s, \action, s') > 0$ implies $\prob'(s, \action, s') > 0$.
An \emph{end-component} (EC) of an MDP $\M$ is a sub-MDP $\M'$ of $\M$ such that its underlying graph is strongly connected.
A maximal end-component (MEC) is an EC $\E = (E, \act', \alphabet, \prob', \lab)$ such that there is no other EC $\E = (E', \act'', \alphabet, \prob'', \lab)$ such that $E \subset E'$.
An MEC $\E$ that cannot reach states outside $\E$ is a \emph{leaf} component.

\begin{theorem}[\cite{DBLP:phd/us/Alfaro97,DBLP:books/daglib/0020348}]\label{thm:end-component}
    Once an EC $\E$ of an MDP is entered, there is a strategy that visits every state-action combination in $\E$ with probability $1$ and stays in $\E$ forever. Moreover, for every strategy the union of the end-components is visited with probability $1$.
    An infinite path of an MC $\M$ almost surely (with probability $1$) will enter a leaf component.
\end{theorem}

\paragraph{Linear Temporal Logic.}

An \ltl formula, over a finite set of atomic propositions $\ap$ is defined as $\varphi ::=   a \in \ap \mid \neg \varphi \mid \varphi \land \varphi \mid \varphi \lor \varphi \mid \ltlfX \varphi \mid \ltlfG\varphi \mid \ltlfF\varphi \mid \varphi \ltlfU \varphi$.
Here $\ltlfX$ (Next), $\ltlfG$ (Globally/Always), $\ltlfF$ (Finally/Eventually) and $\ltlfU$ (Until) are temporal operators.

An $\omega$-trace $w$ is an infinite sequence of letters $w[0]\,w[1]\,w[2]\dots$ with $w[i] \in \alphabet = 2^{\ap}$. We denote the infinite suffix $w[i]\,w[i+1]\dots$ by $w_i$. The satisfaction relation $\models$ between $\omega$-traces $w$ and formulas $\varphi$ is defined as follows:

\[
\begin{array}{ll}
w \models a                     \!& \iff a \in w[0], \\
w \models \lnot \varphi         \!& \iff w \not\models \varphi, \\
w \models \ltlfX\varphi         \!& \iff w_1 \models \varphi, \\
w \models \ltlfG\varphi         \!& \iff \forall k.\; w_k \models \varphi, \\
w \models \ltlfF\varphi         \!& \iff \exists k.\; w_k \models \varphi, \\
w \models \varphi \ltlfU \psi   \!& \iff \exists k.\bigl(w_k \models \psi \land \forall\, j < k,\; w_j \models \varphi\bigr) \\
\end{array} 
\]
Further, $\forall w. w \models \ttrue$ and $\forall w. w \not \models \ffalse$.
We denote by $\flang{\varphi}$ the set of \emph{infinite traces} satisfying the \ltl formula $\varphi$.

\paragraph{Automata.}

A (nondeterministic) transition system (TS) is a tuple $\T = (\states, \init, \trans)$, with a finite set of states $\states$, an initial state $\init \in \states$, and a transition function $\trans: \states \times \alphabet \rightarrow 2^{\states}$. We extend $\trans$ to sets by $\trans (S, \letter) := \bigcup_{q\in S}\trans(q, \letter)$.

A \emph{deterministic} TS satisfies $\size{\trans(q, \letter)} \leq 1$ for each $q \in \states$ and $\letter \in \alphabet$.
An automaton $\A$ is defined as a tuple $(\T, \acccond)$, where $\T$ is a TS and $\acccond$ is an acceptance condition.

We define $\Delta(\trans) = \{(q, \letter, q') \mid q, q'\in \states, \letter \in \alphabet, q' \in \delta(q, \letter)\}$.

For an infinite trace $w\in\alphabet^\omega$, a \emph{run} of $\A$ on $w$ is an infinite sequence of transitions $\run = (q_{0},w[0], q_1)(q_1, w[1], q_2)\cdots$ such that, for every $i \geq 0$, $q_{i+1} \in \trans(q_{i}, w[i])$.
Let $\inf(\run)$ be the set of transitions that occur infinitely often in a run $\run$.
The acceptance condition $\alpha \subseteq \Delta(\delta)$ is a set of \emph{accepting} (\emph{rejecting}, resp.) transitions for \buchi (co-\buchi, resp.). A run $\run$ satisfies the \buchi (co-\buchi, resp.) acceptance condition $\acccond$ if $\inf(\run) \cap \alpha \neq \emptyset$ ($\inf(\run) \cap \alpha = \emptyset$, resp.).

A run is \emph{accepting} if it satisfies the condition $\acccond$;
a trace $w \in \infwords$ is \emph{accepted} by $\A$ if there is an accepting run $\run$ of $\A$ over $w$.
We use three letter acronyms in $\{D, N\} \times \setnocond{B, C} \times \setnocond{A}$ to denote automata types where the first letter stands for the TS mode, the second for the acceptance type and the third for automaton.
For example, DBA means deterministic \buchi automaton.
We assume that all automata are \emph{complete}, i.e., for each state $s\in\states$ and letter $\letter\in\alphabet$, $|\trans(s,\letter)|\geq 1$.

We denote by $\inflang{\A}$ the \emph{$\omega$-language} recognised by an $\omega$-automaton $\A$, i.e., the set of $\omega$-traces accepted by $\A$.

\section{Stochastic Planning Problem}\label{sec:plan-problem}
\paragraph{Overview.}
In a stochastic planning problem, we model the interaction between the agent and the environment as an MDP $\M$ and its task specification as an \ltl formula $\varphi$.
The goal is to find an optimal strategy $\strategy$ for $\M$ that maximises the chance that the infinite sequence of observed propositions satisfies $\varphi$. Formally, we want the probability of the $\omega$-traces generated by $\M^{\strategy}$ and belonging to $[\varphi]$, to be maximal among all possible choices of $\strategy$.
\paragraph{MDPs and LTL Task Specification.}
Every LTL formula $\varphi$ can be translated into a (nondeterministic) \buchi automaton $\A$, accepting $[\varphi]$, i.e., $\lang{\A} = [\varphi]$~\cite{DBLP:conf/lics/VardiW86}.
Recall that for a given strategy $\strategy$ on an MDP $\M$, we can induce an MC $\M^{\strategy}$.
We define the semantic satisfaction probability of the induced MC $\M^{\strategy}$ for $\lang{\A}$ as 
$\pp_{\M^{\strategy}}(\lang{\A}) = \pp\setnocond{ \xi \in \Omega_\strategy^{\M}(s_0): \lab(\xi) \in \lang{\A}}$, which is the probability of $\M^{\strategy}$ generating an $\omega$-trace in $[\varphi]$.
For an MDP $\M$ and an automaton $\A$, we define the \emph{maximal semantic satisfaction probability} as
$\psem(\M, \A) = \sup_{\strategy} \pp_{\M^{\strategy}}(\lang{\A})$.
The strategy $\strategy$ that achieves $\psem(\M, \A)$ is the \emph{optimal} strategy we look for in the planning problem.

\paragraph{Product of MDPs and \buchi automata.} If $\A$ is a DBA, we can solve the problem of finding an optimal strategy $\strategy$ by constructing the product MDP $\M \times \A$ and extract a memoryless strategy $\strategy$ on $\M\times \A$ that reaches accepting components with maximal probability~\cite{DBLP:books/daglib/0020348}.
If $\A$ is nondeterministic, this is in general not possible, because we have to resolve the nondeterminism, which corresponds to future forecasting capabilities that we can assign neither to the agent, nor to the (probabilistic) environment.
However, it has been observed ~\cite{DBLP:conf/tacas/HahnPSS0W20,DBLP:conf/cav/SickertEJK16,DBLP:conf/concur/HahnLST015} that there is a class of automata called good-for-MDPs (GFM) for which we can assign the nondeterminism to the agent without loss of optimality. Intuitively, a nondeterministic automaton is GFM, if the nondeterminism can be resolved by the agent (we postpone the formal definition of GFM to later in this section). If an automaton is GFM, then we can still adopt a variant of the product construction shown below:

Formally, let $\M = (S, \act, \alphabet, \prob, s_0, \lab)$ be an MDP and $\A = (\states, q_0, \trans, \acccond)$ be an NBA.
We define the product MDP $\M \times \A = (S^{ \times}, \act^{\times}, \alphabet, \prob^{\times}, \langle s_0, q_0\rangle, \lab^{\times}, \acccond^{\times})$ augmented with the acceptance condition $\acccond^{\times}$ where

\begin{itemize}\itemsep=0pt
    \item $S^{\times} = S \times Q$ is the state space.
    \item $\act^{\times }= \act \times [k]$ is the action set where $k$ is the \emph{maximal} out-degree of $\A$ for any state in $\states$ and letter in $\alphabet$. We denote $[k] = \{1, \cdots, k\}$ for any integer $k > 0$.
    
    \item $\prob^{\times}: S^{\times} \times \act^{\times} \times S^{\times} \rightarrow [0,1]$ is the transition probability function such that $\prob^{\times}(\langle s, q\rangle, \langle \action, i\rangle, \langle s', q'\rangle) = \prob(s, \action, s')$ if $\prob(s, \action, s') > 0$ and $q' \in \trans(q, \lab(s, \action))$ where $q'$ is the $i$-th state in the \emph{ordered} set $\trans(q,  \lab(s, \action))$.
    \item $\lab^{\times}(\langle s, q\rangle, \langle \action, i\rangle ) = \lab(s, \action)$ for state $\langle s, q\rangle \in S \times \states$ and action $\langle \action, i\rangle \in \act^{\times}$, and
    \item For \buchi/co-\buchi, $\acccond^{\times} = \setnocond{(\langle s, q\rangle, \langle \action, i\rangle, \langle s',q'\rangle )  \mid \prob^{\times}(\langle s, q\rangle, \langle \action, i\rangle, \langle s',q'\rangle ) > 0, (q, \lab(s,\action), q') \in \acccond}$.
\end{itemize} 
Instead of storing the selected successor state in the action name as in other literature, e.g.~\cite{DBLP:conf/tacas/HahnPSS0W20}, we index the choice by its position in a predefined order over successors.
Formally, for each $q \in \states$ and $\letter \in \alphabet$, we define a function $\order_{\A, q,\letter}$ mapping each $q' \in \trans(q,\letter)$ to its unique position $i \in [k]$.
This index $i$ serves as the identifier for the agent’s choice in our product construction.
This encoding is equivalent to ~\cite{DBLP:conf/tacas/HahnPSS0W20} but has a slight advantage of using fewer actions. For full details, see Appendix~\ref{appendix:action-indexing}.

\paragraph{Conditions for GFMness.}
In order to see if an NBA is GFM, we have to characterise the probabilities that come from the syntactic construction above and show that they are the same of those intrinsic in the original problem. 
Recall that the probability of the original problem is the following:
\[
\psem(\M, \A) = \sup_{\strategy} \pp_{\M^{\strategy}}(\lang{\A})
\]
The probability coming from the syntactic product construction above is characterised as:
\[\psyn(\M, \A) = \sup_{\strategy} \pp \setnocond{\xi \in \Omega^{\M\times\A}_{\strategy}(\langle s_0,q_0\rangle) : \xi \text{ accepting}}.\] This can be simplified to
$\psyn(\M, \A) = \sup_{\strategy}\pp_{(\M\times\A)^{\strategy}}(\ltlfF X)$, where $X$ is the set of states of the accepting MECs in $\M\times\A$ that contain accepting transitions in $\acccond^{\times}$.

Clearly, $\psyn(\M, \A) \leq \psem(\M,\A)$, because accep\-ting runs $\xi$ only occur on accepting words.
Thus, a strategy $\strategy$ chooses an accepting run on accepting words in the best case, but it is also possible for $\strategy$ to make wrong decisions.

Obviously, $\psyn(\M, \A) = \psem(\M,\A)$ if $\A$ is deterministic. 
Following \cite{DBLP:conf/tacas/HahnPSS0W20}, an NBA $\A$ is said to be GFM if, for all finite MDPs $\M$, $\psem(\M, \A) = \psyn(\M,\A)$ holds. See Appendix \ref{app:non-gfm} for a non-GFM case.

\paragraph{LTL to GFM automata.}
There are several algorithms to transform an LTL formula into an NBA that is GFM~\cite{DBLP:conf/cav/SickertEJK16, DBLP:conf/tacas/HahnPSS0W20}. 
The runtime to construct a GFM automaton from an LTL formula is doubly exponential in the size of the formula, which is the same cost as obtaining a deterministic automata.
However, the practical advantage of going through GFM instead of deterministic automata is enormous and all state-of-the-art systems implement an LTL to GFM construction \cite{DBLP:conf/cav/SickertEJK16, DBLP:conf/tacas/HahnPSS0W20}.

\section{Probabilistic $\omega$-Automata}
Our state reduction technique is based on obtaining a small probabilistic (\buchi) automaton (PA) out of a GFM automaton.
The key characteristic of PAs \cite{DBLP:journals/jacm/BaierGB12} is that, while being nondeterministic, their nondeterministic choices are resolved randomly. A PA $\pa = (\alphabet, \states, \trans, q_0, \acccond)$ is a nondeterministic automaton equipped with a \buchi acceptance condition $\acccond$ and a randomised transition function $\delta : Q \times \alphabet \mapsto \distr(Q)$, where, from state $q$ and letter $\letter$, transition to $q'$ is taken with probability $\delta(q,\letter)(q')$. 
We often abuse notation by writing $\delta(q,\letter)$ to denote its support.
Each word $w \in \infwords$ induces a probability measure $\pp^w_{\pa}$ on $\states^{\omega}$ in the usual way. 
The probability that $\pa$ accepts $w$, denoted by $\pp_{\pa}(w)$, is the probability measure of all accepting runs of $w$ on $\pa$, that is: \[\pp_{\pa}(w) = \pp^w_{\pa}(\{ \pi: \pi \text{ is an accepting run of } w \}).\]

A PA $\pa$ is a 0/1-PA if, for any word $w \in \alphabet^{\omega}$, we have either $\pp_{\pa}(w) = 1$ or $\pp_{\pa}(w) = 0$.
For a 0/1-PA $\pa$, with a slight abuse of notation, we say a word $w$ is accepted by $\pa$ ($w$ is in $\lang{\pa}$) if $\pp_{\pa}(w) = 1$.
0/1-PAs are known to be \emph{semantically deterministic}~\cite{LiPST25}.
That is, given a state $p$ and a letter $\letter$ of 0/1-PA $\pa$, for any two states $q, r\in \delta(p, \letter)$, we have that $\lang{\pa^{q}} = \lang{\pa^r}$ where $\pa^s$ is the automaton by setting the initial state to $s$.
For a detailed introduction to PAs please refer to \cite{DBLP:journals/jacm/BaierGB12}.

\paragraph{Product of MDPs and 0/1-PA.} 
In the following, we show for the \emph{first} time that 0/1-PAs can also be perceived as another type of GFM automata.
To this end, we first define the product MDP of an MDP $\M$ and a 0/1-PA $\pa$
\footnote{We note that a source of 0/1-PAs can be a special type of automata whose nondeterminism can be resolved by randomness discussed in~\cite{henzinger2025resolvingnondeterminismrandomness}.
This type of automata is also said to be GFM in a discussion without formal proofs in~\cite{henzinger2025resolvingnondeterminismrandomness}.}.

Formally we define the product as follows. 
    Let $\M = (S, \act, \alphabet, \prob, s_0, \lab)$ be an MDP and $\pa = (\states, q_0, \delta, \acccond)$ be the 0/1-PA over $\alphabet$.
    We define the product MDP $\M\otimes \pa = (S^{\otimes}, \act^{\otimes}, 
    \alphabet,
    \prob^{\otimes}, \langle s_0, q_0\rangle,
    \lab^{\otimes},\acccond^{\otimes})$ where
    \begin{itemize}
        \item $S^{\otimes} = S \times \states $ is the state space,
        \item $\act^{\otimes} = \act$ is the action set,
        \item $\prob^{\otimes}: S^{\otimes}\times \act^{\otimes} \times S^{\otimes} \rightarrow [0, 1]$ is the transition probability function such that $\prob^{\otimes}(\langle s, q\rangle,  \action, \langle s', q'\rangle) = \prob(s, \action, s') \cdot \trans(q, \letter, q')$ if $\prob(s, \action, s') > 0$ and $q' \in \trans(q, \letter)$ where $\letter = \lab(s, \action)$, 
        \item $\lab^{\otimes}: S^{\otimes} \times \act^{\otimes} \rightarrow \alphabet$ where $\lab^{\otimes}(\langle s, q\rangle, a) = \lab(s, a)$ and

        \item $\acccond^{\otimes} =\{(\langle s,q\rangle,  \action, \langle s',q'\rangle ) \mid (q, L(s, a), q') \in \acccond,(q, \letter, q') \in \acccond,  \action\in \act^{\otimes}, \prob^{\otimes}(\langle s, q\rangle, \action, \langle s',q'\rangle) > 0
        \}$.
    \end{itemize}

Intuitively, we still ask the agent to make decisions on which letter to choose for the 0/1-PA $\pa$, but leave the choice of a successor to a random strategy.
That is, the agent resolves the nondeterminism by choosing actions, and once the action $a$ is selected, according to the definition of MDP $\M\otimes\pa$,  we also know the chosen letter $\letter = \lab^{\otimes}(\langle s, q\rangle, a)$ as well.
This is different from the classical product $\M \times \A$, where the agent not only decides the next letter, but also the next successor in $\A$, by selecting the action $\langle a, i\rangle \in \act^{\times}$.

For the product $\M\otimes \pa$, we can define a similar maximal \emph{syntactic} satisfaction probability:
\[\psyn(\M, \pa) = \sup_{\strategy} \pp \setnocond{\xi \in \Omega^{\M\otimes\pa}_{\strategy}(\langle s_0,q_0\rangle) : \xi \text{ is accepting}} .\]
Crucially, we can see that 0/1-PAs are another type of GFM \buchi automata in the following sense:
\begin{theorem}\label{thm:pa-gfm}
    Let $\mathcal{D}$ be a DBA and $\pa$ be its equivalent 0/1-PA such that $\lang{\pa} = \lang{\mathcal{D}}$.
    For a finite MDP $\M$, we have that $\psyn(\M, \pa) =  \psyn(\M, \mathcal{D}) = \psem(\M, \mathcal{D}) = \psem(\M, \pa)$.
\end{theorem}

An equivalent 0/1-PA $\pa$ always exists for a DBA $\D$ because a DBA can be easily transformed to an equivalent 0/1-PA by transitioning to the only successor with probability one.
Note that we only consider the \buchi condition in Theorem~\ref{thm:pa-gfm}, but the results can easily be generalised to Rabin conditions.
Since 0/1-PAs are GFM based on our product MDP definition, we can then apply standard strategy synthesis approach once the product MDP $\M\otimes \pa$ is constructed.

\section{Good-for-MDP State Reduction}
\begin{figure}[th]
  \centering
  \includegraphics[width=\linewidth,page=1]{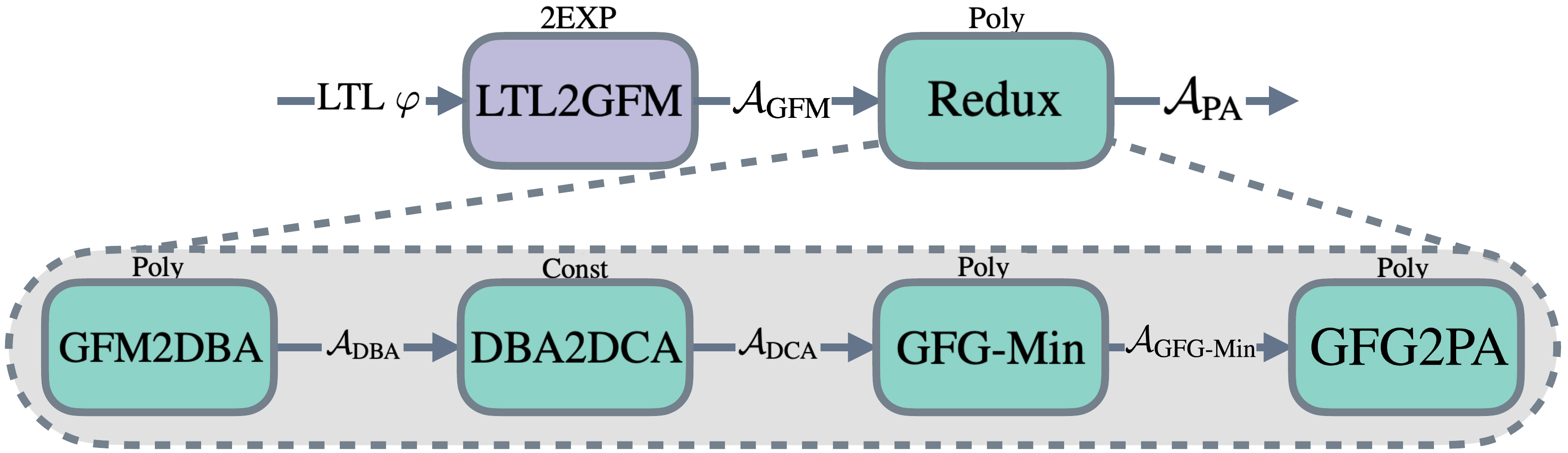}
  \caption{Overview of our state reduction pipeline}
  \label{fig:overview-gfm-reduction}
\end{figure}
\noindent
In this section, we present our main contribution, a general state space reduction for GFM automata. We assume to have a GFM automaton $\Agfm$ and through a polynomial \emph{Redux} procedure,  we return a 0/1-PA $\Apa$, whose state space can be significantly reduced. 
\emph{Redux} works in several stages, see Figure \ref{fig:overview-gfm-reduction}, which we detail below.

 \begin{itemize}
     \item  First step (Poly): we perform a GFM-to-DBA transformation by embedding in the transitions of the GFM $\Agfm$ the choices $[k]$ used in the product construction of Section~\ref{sec:plan-problem}, making it a complete DBA $\Adba$.
     \item  Second step (Const): we formally treat the DBA $\Adba$ as a DCA, $\Adca$, by simply changing the acceptance condition.
     \item  Third step (Poly): we apply a minimisation algorithm originally developed for good-for-games (GFG) NCAs \cite{AbuRadiK22} to $\Adca$ to obtain a minimal GFG-NCA $\Agfgmin$.
     \item  Fourth step (Poly): we transform the GFG-NCA  $\Agfgmin$ into a 0/1-PA $\Apa$ with the same number of states as $\Agfgmin$.
 \end{itemize}
Here ``Poly'' and ``Const'' indicate that their corresponding steps perform in polynomial and constant time, respectively. 
We now detail each step below.

\subsection{Step 1: GFM to DBA}
We consider the product $\M \times \Agfm$ of Section 3 and modify both the MDP $\M$ and automaton $\Agfm$ by embedding the transition choices $[k]$. 
The resulting MDP $\M'$ has its action set expanded from $\act$  to  $\act \times [k]$.
Similarly, the resulting automaton $\Adba$ will have the transitions re-labelled from $\Sigma$ to $\Sigma \times [k]$. As a result the automaton $\Adba$ is deterministic. 
Formally, we define the modified MDP $\M' = (S', s'_0, \act', \prob', \lab')$ from $\M = (S, s_0, \act, \prob, \lab)$ 
as 
\begin{itemize}
    \item $S' = S$, $s'_0 = s_0 $, $\act' = \act \times [k]$,
    \item $\lab': S' \times \act' \rightarrow \alphabet'$ where $\alphabet' = \alphabet \times [k]$, and if $\lab(s, \action) = \letter$, then $\lab'(s, \langle \action, i\rangle) = \langle \letter, i\rangle$ for all $i \in [k]$, and 
    \item $\prob'(s, \langle \action, i\rangle, s') = \prob(s, \action, s')$ for all $i \in [k]$ and $\prob(s, \action, s') > 0$.
\end{itemize}

We define the DBA $\Adba = (\states_{\text{DBA}}, q_{0}, \trans_{\text{DBA}}, \acccond_{\text{DBA}})$ from $\Agfm = (\states, q_0, \trans, \acccond)$ as:
\begin{itemize}
    \item $\states_{\text{DBA}} = \states$, 
    $\acccond_{\text{DBA}} = \{(q, \langle \letter, i\rangle, q') \mid (q, \letter, q')\in \acccond, i = \order_{\A, q, \letter}(q')\}$, and
    \item $\trans_{\text{DBA}}(q, \langle \letter, i\rangle) = q'$ if $q' \in \trans(q, \letter)$ and $\order_{\A,q, \letter}(q')= i$.
\end{itemize}
$\Adba$ will then be made complete.
Intuitively, we assign the nondeterminism of $\Agfm$ to the agent by extending the action set $\act$ to $\act\times [k]$ in $\M'$, which then would allow us to obtain the DBA $\Adba$ since the nondeterminism in $\Agfm$ will be taken care of by the agent.
We observe that $\M\times \Agfm$ and $\M'\times \Adba$ (obtained by applying the product construction in Section 3) have the same state space.
Moreover, in $\M' \times \Adba$, the action set is $\act \times [k]\times [1]$, as the maximal out-degree of the DBA $\Adba$ is $1$. Notice that we can drop $[1]$, which is a constant in every transition and if we do so, we get the following:
\begin{lemma}\label{lem:syn-same-prob}
The product $\M \times \Agfm$ and $\M'\times\Adba$ are the same.
Moreover, the optimal strategy $\strategy$ for each state pair $(s, q) \in S\times Q$ in either of the products will obtain optimal satisfaction probabilities in both products.    
\end{lemma}

By Lemma~\ref{lem:syn-same-prob}, it is easy to observe that the syntactic probabilities of $\M \times \Agfm$ and $\M'\times \Adba$ are the same.
Since $\Adba$ is deterministic and thus GFM, we have $\psyn(\M', \Adba) = \psem(\M', \Adba)$.
It then follows:
\begin{theorem}\label{thm:equivalent-max-prob}
    $\psem(\M,\Agfm) = \psyn(\M, \Agfm) = \psyn(\M', \Adba) = \psem(\M', \Adba)$.
\end{theorem}

The GFM to DBA conversion can be done in polynomial time. The construction of the modified MDP $\M'$ and the DBA $\Adba$ involves a direct translation of the original components. The number of states remains the same, while the size of the action set and the alphabet expands by a factor of $k$, which is a polynomial increase.

\subsection{Step 2: DBA to DCA}
Next, we syntactically convert the DBA $\Adba$ into a DCA $\Adca$.
To do so, we only need to modify the acceptance condition. $\Adca$ has the same set of states, initial state, and transition function as $\Adba$. The co-\buchi acceptance condition requires that a run is accepting if it intersects with the set of rejecting transitions only a finite number of times. The set $\acccond_{\text{DCA}}$ of rejecting transitions for the DCA $\Adca$ is the same as the set $\acccond_{\text{DBA}}$ of accepting transitions in $\Adba$, but only interpreted differently. Observe that the languages of $\Adba$ and $\Adca$ complement each other. $\Adca$ is just an intermediate automaton on which we apply GFG-minimisation next.

Since this conversion does not require any computational modification to the automaton's transition structure, it can be done in constant runtime.

\subsection{Step 3: GFG-Minimisation}
$\Adca$ is deterministic and thus a GFG automaton that we can apply minimisation algorithms~\cite{AbuRadiK22} on in this step. To formally understand this step, we take a detour to introduce GFG automata.

 GFG automata \cite{DBLP:conf/csl/HenzingerP06} are automata in which nondeterminism can be resolved \emph{deterministically} by assigning the choice to the protagonist in an adversarial (vs.\ stochastic MDPs) games. 
Formally, an automaton $\A$ is said to be GFG if there exists a strategy $f : \Sigma^* \rightarrow \states$ such that for every accepting word $w = \letter_0 \letter_1\cdots$, the run $f(\epsilon)f(\letter_0) \cdots f(\letter_0\cdots \letter_i) \cdots $ is accepting.
Intuitively, this indicates that there is a deterministic strategy to produce an accepting run in a GFG automaton even if the input accepting word is given letter by letter.
Interestingly, GFG
 co-\buchi automata can be minimised in \emph{polynomial} time~\cite{AbuRadiK22}.
Since GFG automata have more restricted nondeterminism than GFM automata, GFG automata are also GFM~\cite{DBLP:conf/lata/KleinMBK14}.
GFG automata are not adopted in state-of-the-art stochastic planning tools, such as PRISM~\cite{Prism40}.
The reason is that GFG \buchi (resp.\ co-\buchi) automata cannot recognise all languages expressed by \ltl as they are only as expressive as their deterministic counterparts.
Moreover, current approaches to construct GFG automata are not as efficient as the ones for GFM.
Notably, our work is able to use GFG minimisation on GFM automata that recognise all $\omega$-regular properties. See Appendix~\ref{app:omega-reg-preservation} for more details.

In our construction we apply GFG co-\buchi minimisation~\cite{AbuRadiK22} on $\Adca$, yielding a minimal GFG-NCA $\Agfgmin$ accepting $\lang{\Adca}$.

\subsection{Step 4: GFG-Min to 0/1-PA}

\cite{LiPST25} has proven that the minimal GFG-NCA produced by~\cite{AbuRadiK22} can be turned into a language equivalent 0/1-PA by resolving its nondeterminism with random choices.

The translation from the minimal GFG-NCA $\Agfgmin$ to a 0/1-PA $\Apa$ is quite simple:
we only need to resolve the nondeterminism by random choices.
Formally, we treat $\Agfgmin$  as a \buchi automaton $\Anba = (\states_{\NBA}, q_0, \trans_{\NBA}, \acccond_{\NBA})$ and build the PA $\Apa = (\states_{\text{PA}}, q_0, \trans_{\text{PA}}, \acccond_{\text{PA}})$ as:
\begin{itemize}
    \item $\states_{\text{PA}} = \states_{\NBA}, \acccond_{\text{PA}} =\acccond_{\NBA}$, and
    \item $\trans_{\text{PA}}(q, \letter)(q') = \frac{1}{|\trans_{\NBA}(q, \letter)|}$ for each $q \in \states_{\text{PA}}, \letter \in \alphabet'$ and $q' \in \trans_{\NBA}(q, \letter)$.
\end{itemize}

Following from~\cite[Lemma 3]{LiPST25}, we get:
\begin{lemma}\label{lem:pa-lang}
    $\lang{\Apa} = \lang{\Adba}$ and $\Apa$ is a 0/1-PA. 
\end{lemma}
It immediately follows that, according to Theorem~\ref{thm:pa-gfm}, we can use $\Apa$ to obtain an optimal strategy for $\M'$ to achieve $\psem(\M', \Apa) = \psem(\M', \Adba)$ since $\lang{\Apa} = \lang{\Adba}$ holds. 
Together with Theorems~\ref{thm:pa-gfm} and~\ref{thm:equivalent-max-prob}, we then obtain our main result:
\begin{theorem}\label{thm:pa-main}
From an optimal strategy $\strategy$ on $\M'\otimes \Apa$, one can obtain an optimal strategy $\strategy'$ for $\M$ that achieves the maximal probability $\psem(\M, \Agfm)$. 
\end{theorem}

\paragraph{Summary.} Details on optimal strategy synthesis, along with further optimisations, are in Appendix \ref{appendix:optimal-strategy-further-optimisations}.
In summary, our stochastic planning algorithm can be formalised below: 

\begin{itemize}
\item Construct a GFM (\buchi) automaton $\Agfm$ from $\varphi$.
\item Create $\M'$ from $\M$ and  $\Adba$ (and thus $\Adca$) from $\Agfm$.
\item Apply GFG minimisation on $\Adca$ and treat the minimised automaton $\Agfgmin$ as a \buchi automaton $\Anba$.
\item Translate $\Anba$ to a 0/1-PA $\Apa$.
\item Create $\M' \otimes \Apa$.
\item Identify the accepting MECs that have some accepting transitions.
\item Synthesise an optimal strategy $\strategy$ that maximises the reachability probability of accepting MECs as usual. 
\end{itemize}

\section{Direct GFM Construction for $\ltlfG\ltlfF \varphi$}
\paragraph{$\ltlfG\ltlfF \varphi$ fragment.}
We now focus on the syntactic class of LTL formulas of the form $\ltlfG\ltlfF \varphi$ where $\varphi$ 
is a \emph{co-safety} formula with the following syntax:
\begin{align*}
\varphi := a \mid \neg a \mid \varphi \land \varphi \mid \varphi \lor \varphi \mid \ltlfX \varphi \mid \ltlfF \varphi \mid \varphi \ltlfU \varphi.
\end{align*}
They express that a finite trace pattern repeats infinitely often, which we refer to as \emph{repeated reachability} properties.
Formally, $P\subseteq \alphabet^\omega$ is a repeated reachability property if there exists a language of
finite words $R \subseteq \alphabet^*$ such that, for every $w \in P$, infinitely
many prefixes of $w$ belong to $\finwords\cdot R$.
Then, we have that:
\begin{lemma}
\label{reapeated-reachability}
    $P$ is a repeated reachability property specifiable in LTL if, and only if, $P$ is specifiable in $\ltlfG\ltlfF \varphi$, where $\varphi$ is a co-safety formula.
\end{lemma}

The $\ltlfG\ltlfF \varphi$ fragment is very common in verification \cite{holevcek2004verification} and also in reinforcement learning \cite{deepltl}. More generally, repeated reachability properties are within a commonly used subclass of so-called \emph{recurrence properties} \cite{DBLP:conf/podc/MannaP89,DBLP:conf/icalp/ChangMP92}.

\paragraph{Direct GFM Construction for $\ltlfG\ltlfF \varphi$.}
\label{subsection:GFM_gf}
Consider a formula of the form $\ltlfG\ltlfF \varphi$, the subformula $\varphi$ expresses a \emph{finite trace property}, which can be accepted by a nondeterministic finite automaton (NFA)\footnote{Our construction applies to any NFA, not just those from co-safety LTL (see Appendix \ref{app:co-safety-need}), covering repeated reachability properties not expressible in LTL, i.e., monadic first-order and finite LTL~\cite{de2013linear}.}.
To make our construction more general, we will start from an NFA instead of the formula $\varphi$ in the following.
An NFA $\N $ is a tuple $(\T, F)$ where $\T=(\states, q_0, \trans)$ is a \emph{nondeterministic} TS and $F \subseteq \states$ is a set of \emph{final} states.
Unlike NBAs, NFAs only operate on \emph{finite} traces rather than $\omega$-traces.
A finite trace $u \in \finwords$ is accepted by an NFA $\N$ if one of its finite runs \emph{terminates} at a final state.

We can construct for a co-safety property formula $\varphi$, an NFA $\N_{\varphi} = (\states, q_0, \trans, F)$ such that $\finlang{\N_{\varphi}}\cdot (\ttrue)^{\omega} = [\varphi]$ where $\finlang{\N}$ denotes the set of finite traces accepted by ${\N_{\varphi}}$.
Note that the alphabet here is $\alphabet = 2^{\ap}$.

Now we can construct from $\N_{\varphi}$ a GFM automaton $\A = ( \states_{\A}, q_{\A}, \trans_{\A}, \acccond_{\A})$ for $\ltlfG\ltlfF \varphi$ where 
\begin{itemize}
\item $\states_{\A} \subseteq \left((\states \setminus F) \cup \{q_0\}\right)$, $q_{\A} = q_0$, 
\item $\trans_{\A}: \states_{\A} \times \alphabet \rightarrow 2^{\states_{\A}}$ is defined such that for each $q \in \states_{\A}$ and $\letter \in \alphabet$, we have 
(1) $q' \in \trans_{\A}(q,\letter)$ if $q' \in \trans(q, \letter)$ with $q' \notin F$, and (2) $q_0 \in \trans_{\A}(q, \letter)$, and
\item 
 $\acccond_{\A} = \{(q,\letter,q_0) \in \Delta(\trans_{\A}) \mid \exists q' \in \trans(q, \letter) \land q' \in F\}$.
\end{itemize}
The intuition is that, for a repeated reachability property $\ltlfG\ltlfF \varphi$, we can forget the past finite trace at \emph{any} point and start tracking whether the following finite trace satisfies $\varphi$.
The \buchi acceptance condition will make sure that the reachability/co-safety formula $\varphi$ will be satisfied infinitely often.
Therefore, in the construction, we can always reset and go back to the initial state $q_0$, which is why every state $q$ has a successor $q_0$ on every letter $\letter \in \alphabet$.
Once the formula $\varphi$ has been fulfilled, i.e., a final state $q'$ has been reached from $q$ over letter $\letter$ in $\N_{\varphi}$, we need to mark the transition $(q, \letter, q_0)$ in $\A$ as accepting and start tracking finite traces satisfying $\varphi$ again by moving back to the initial state $q_0$.

It is also easy to show that a memoryless random strategy on $\A$ can generate an accepting run almost surely over an $\omega$-trace from $[\ltlfG\ltlfF\varphi]$.
This is because accepting transitions can be reached with positive probabilities from anywhere in $\A$ and it is not possible to skip accepting transitions forever in an infinite run with positive probabilities.

Let $|\varphi|$ be the number of modalities and connectives in $\varphi$.
It then follows that:
\begin{theorem}\label{thm:GF-fragment-construction}
    (1) $\lang{\A} = [\ltlfG\ltlfF\varphi]$, (2) $\A$ is GFM and (3) $\A$ has ${2^{\mathcal{O}(|\varphi|)}}$ states.
\end{theorem}

Note that as, for a co-safety/reachability property $\varphi$, the NFA $\N_{\varphi}$ of $\varphi$ has $2^{\mathcal{O}(|\varphi|)}$ states~\cite{de2013linear}, so does our GFM automaton $\A$ for $\ltlfG\ltlfF \varphi$.
Current GFM constructions such as~\cite{DBLP:conf/cav/SickertEJK16,DBLP:conf/tacas/HahnPSS0W20} normally output a DBA for $\ltlfG\ltlfF \varphi$, which in general has $2^{2^{\mathcal{O}(|\varphi|)}}$ states.
Therefore, our specialised GFM construction for repeated reachability property $\ltlfG\ltlfF \varphi$ can be \emph{exponentially more efficient} than current approaches.

\begin{table}[t]
\centering
\footnotesize
\setlength{\tabcolsep}{3pt}
\renewcommand{\arraystretch}{1.2}
\begin{tabular}{
    >{\centering\arraybackslash}m{1cm}  
    >{\centering\arraybackslash}p{1.5cm}  
    >{\centering\arraybackslash}p{1.55cm}  
    >{\centering\arraybackslash}p{1.5cm}  
    >{\centering\arraybackslash}p{1.65cm}  
}
\toprule

\textbf{Pattern} & \texttt{Owl} & \texttt{Owl-Red} & \texttt{Slim} & \texttt{Slim-Red} \\
\midrule

{$\text{TDR}[6]$} & 64 ({\scriptsize 0.46}) & 64 ({\scriptsize 0.03}) & 65 ({\scriptsize 0.09}) & \textbf{34} ({\scriptsize 0.01}) \\
{$\text{TDR}[7]$} & 128 ({\scriptsize 0.46}) & 128 ({\scriptsize 0.15}) & 129 ({\scriptsize 0.12}) & \textbf{66} ({\scriptsize 0.02}) \\
{$\text{TDR}[8]$} & 256 ({\scriptsize 0.48}) & 256 ({\scriptsize 0.69}) & 257 ({\scriptsize 0.14}) & \textbf{130} ({\scriptsize 0.05}) \\
\midrule

{$\text{LIB}[4]$} & 17 ({\scriptsize 0.44}) & \textbf{10} ({\scriptsize 0.01}) & 33 ({\scriptsize 0.11}) & 18 ({\scriptsize 0.13}) \\
{$\text{LIB}[5]$} & 21 ({\scriptsize 0.48}) & \textbf{12} ({\scriptsize 0.01}) & 65 ({\scriptsize 0.20}) & 34 ({\scriptsize 3.83}) \\
{$\text{LIB}[6]$} & 25 ({\scriptsize 0.62}) & \textbf{14} ({\scriptsize 0.01}) & 129 ({\scriptsize 0.69}) & 66 ({\scriptsize 128.34}) \\
\midrule

{$\text{BRP}[6]$}& 69 ({\scriptsize 0.55}) & \textbf{17} ({\scriptsize 0.02}) & 317 ({\scriptsize 0.14}) & 65 ({\scriptsize 0.17}) \\
{$\text{BRP}[7]$} & 133 ({\scriptsize 0.65}) & \textbf{19} ({\scriptsize 0.04}) & 637 ({\scriptsize 0.21}) & 95 ({\scriptsize 0.71}) \\
{$\text{BRP}[8]$} & 261 ({\scriptsize 0.86}) & \textbf{21} ({\scriptsize 0.14}) & 1277 ({\scriptsize 0.33}) & 146 ({\scriptsize 3.49}) \\

\midrule
{$\text{EHP}$} & 13 ({\scriptsize 0.43}) & \textbf{9} ({\scriptsize 0.01}) & 127 ({\scriptsize 0.14}) & 54 ({\scriptsize 0.31}) \\
\midrule
{$\text{NU}[4]$} & 44 ({\scriptsize 1.16}) & \textbf{23} ({\scriptsize 0.02}) & 51 ({\scriptsize 0.22}) & 39 ({\scriptsize 0.02}) \\
{$\text{NU}[5]$} & 150 ({\scriptsize 1.41}) & \textbf{56} ({\scriptsize 0.17}) & 232 ({\scriptsize 0.82}) & 159 ({\scriptsize 0.76}) \\
{$\text{NU}[6]$} & 433 ({\scriptsize 5.92}) & \textbf{249} ({\scriptsize 27.75}) & 1425 ({\scriptsize 12.24}) & 753 ({\scriptsize 163.32}) \\
\midrule

{$\text{LFR}[6]$} & 39 ({\scriptsize 0.77}) & 40 ({\scriptsize 0.27}) & 40 ({\scriptsize 0.44}) & \textbf{21} ({\scriptsize 0.19}) \\
{$\text{LFR}[7]$} & 72 ({\scriptsize 0.92}) & 73 ({\scriptsize 2.99}) & 73 ({\scriptsize 1.57}) & \textbf{34} ({\scriptsize 1.76}) \\
{$\text{LFR}[8]$} & 137 ({\scriptsize 1.45}) & 138 ({\scriptsize 39.11}) & 138 ({\scriptsize 10.74}) & \textbf{59} ({\scriptsize 20.84}) \\
\bottomrule
\end{tabular}
\caption{Comparison of automata state spaces of \texttt{Owl} and \texttt{Slim} against their reduced versions, \texttt{Owl-Red} and \texttt{Slim-Red}. We report number of states (smallest in bold) and runtime (in seconds). Within \texttt{Owl-Red} and \texttt{Slim-Red} we only measure the time it took to reduce the automaton.}
\label{tab:evaluation}
\end{table}

\section{Experiments for GFM state reduction}

In this section, we validate the performance of our GFM state space reduction on LTL specifications extracted from literature, which we define blow.
The full benchmark set, demonstrating consistent state space reduction on complex LTL specifications from over 8 sources, is in Appendix~\ref{appendix:general-gfm-state-space}.
\\
\noindent

{\textbf{Trigger with Delayed Response (TDR).}}
This pattern requires $a$ followed by $b$ after $n$ steps, to hold infinitely often. We define this fragment as 
$\text{TDR}[n] = \ltlfG \ltlfF(a \land \ltlfX^n b)$
where $\ltlfX^n$ denotes $n$ nested applications of the next operator.
\\
\noindent
{\textbf{Liberouter (LIB).}} 
An example from the Liberouter verification project \cite{holevcek2004verification} checks liveness of binary signals, ensuring at least one signal does not become permanently stuck. 
Formally, 
$\text{LIB}[n] = \ltlfG \ltlfF((a \land \ltlfX \neg a) \lor (\neg a \land \ltlfX a) \lor (b \land \ltlfX \neg b) \lor (\neg b \land \ltlfX b) \dots )$, with $n$ numbers of signals. 

\noindent
{\textbf{Bounded Retransmission Protocol (BRP).}} 
Based on \cite{DBLP:journals/jcss/BaierK00023}, BRP models that whenever a message is sent, a corresponding acknowledgement must occur within a bounded number of steps. 
$\text{BRP}[n] = \ltlfG(\texttt{"msg\_sent"} \rightarrow \ltlfF(\texttt{"ack\_send"} \land \varphi_n))$, where the subformula $\varphi_n$ ensures that \texttt{"ack\_rev"} occurs within $n$ steps after \texttt{"ack\_send"}. 
We define $\varphi_n = \texttt{"ack\_rev"} \lor \ltlfX(\texttt{"ack\_rev"} \lor \ltlfX(\dots \lor \ltlfX (\texttt{"ack\_rev"})))$ with $\ltlfX$ applied $n$ times. A smaller $n$ enforces stricter guarantees, while a larger $n$ relaxes them.

\noindent
{\textbf{Etessami–Holzmann Patterns (EHP).}} 
To evaluate the effectiveness of \buchi automata optimisations, \cite{DBLP:conf/concur/EtessamiH00} present deeply nested specifications. One formula is given by 
$\text{EHP} = a \ltlfU(b \land \ltlfX (c \land \ltlfF( d \land \ltlfX \ltlfF( e \land \ltlfX \ltlfF(f \land \ltlfX \ltlfF g)))))$. 
It expresses that condition $a$ must hold continuously until a specific sequence of events is triggered.

\noindent
{\textbf{Nested Until Dependencies (NU).}}
We define the nested-until pattern $\text{NU} [n] = \ltlfG(p_1 \rightarrow \varphi_n)$ where $\varphi_k$ is recursively defined by $\varphi_k = p_k \ltlfU \varphi_{k+1}$ for $k<n$, and $\varphi_n = p_n \ltlfU p_{n+1}$. This pattern was used in the runtime verification benchmarks for SystemC models \cite{DBLP:journals/fmsd/TabakovRV12}.

\noindent
{\textbf{Layered Fairness and Reachability (LFR).}} 
As part of their evaluation, \cite{DBLP:journals/corr/abs-1709-02102} introduced 
$\text{LFR} [n] = \ltlfF(b_1) \land (\ltlfF(b_2 \land \dots \ltlfF(b_n) ) \land \ltlfG \ltlfF(a_1 \land \ltlfX(a_2 \land \dots \ltlfX(a_n)))$. The left-hand conjunct captures a nested reachability, 
while the right-hand side encodes a fairness condition, requiring 
$a_1, \dots, a_n$ to reoccur, with strict temporal progression.
\paragraph{Experiment Setup.}
In order to translate these LTL specifications to GFM automata, we use two state-of-the-art approaches.
Owl \cite{DBLP:conf/atva/KretinskyMS18}, which we denote by \texttt{Owl}, and the ``slim" GFM construction of \cite{DBLP:conf/tacas/HahnPSS0W20}, which we implemented ourselves and denote by \texttt{Slim}.
We refer to reduced Owl automata as \texttt{Owl-Red} and to reduced ``slim" automata as \texttt{Slim-Red}.
We use SPOT \cite{DBLP:conf/atva/Duret-Lutz13} for parts of patterns, for the GFG minimisation and we have used LTL datasets available in the Owl repository. All experiments were run on an 8-core ARM chip, with 16GB of RAM. 

\paragraph{Results.}
Table \ref{tab:evaluation} compares automata states before and after our reduction. We report the LTL to GFM construction time (in seconds) for \texttt{Owl} and \texttt{Slim}
, and reduction time (excluding initial construction) for \texttt{Owl-Red} and \texttt{Slim-Red}.

Across all LTL patterns, the smallest automata (marked in bold) are consistently reduced ones.
\texttt{Owl} features advanced formula simplification and post-processing optimisations and therefore constructs smaller automata than the ``slim" GFM construction.
Since our ``slim" GFM construction is not as optimised as Owl, we can consistently reduce the state space of all ``slim" GFMs and for cases like TDR and LFR, we obtain smaller \texttt{Slim-Red} than \texttt{Owl-Red}.

Note that for LFR, Owl returns incomplete automata, while our reduction returns complete ones with an explicit sink. See Appendix~\ref{app:at-completion} for details.
For the remaining cases NU, EHP, BRP and LIB, the smallest automata are our \texttt{Owl-Red}. More results, can be found in Appendix~\ref{appendix:general-gfm-state-space}.

\begin{table}[t]
\centering
\footnotesize
\setlength{\tabcolsep}{5pt}
\renewcommand{\arraystretch}{1.2}
\begin{tabular}{
    >{\centering\arraybackslash}m{1cm}  
    >{\centering\arraybackslash}p{1.7cm}  
    >{\centering\arraybackslash}p{1.5cm}  
    >{\centering\arraybackslash}p{1.4cm}  
}
\toprule

{\textbf{Pattern}} & \texttt{Owl-Red} & \texttt{Slim-Red} & \texttt{GFM-GF} \\
\midrule

{$\text{TDR}[6]$} & 64 (\scriptsize 0.49) & 34 (\scriptsize 0.10) & \textbf{7} (\scriptsize 0.08) \\
{$\text{TDR}[7]$} & 128 (\scriptsize 0.61) & 66 (\scriptsize 0.14) & \textbf{8} (\scriptsize 0.08) \\
{$\text{TDR}[8]$} & 256 (\scriptsize 1.17) & 130 (\scriptsize 0.19) & \textbf{9} (\scriptsize 0.08) \\
{$\text{TDR}[9]$} & 512 (\scriptsize 4.27) & 258 (\scriptsize 0.48) & \textbf{10} (\scriptsize 0.10) \\
{$\text{TDR}[10]$} & 1024 (\scriptsize 23.67) & 514 (\scriptsize 0.91) & \textbf{11} (\scriptsize 0.09) \\

\midrule

{$\text{LIB}[6]$} & 14 (\scriptsize 0.63) & 66 (\scriptsize 129.03) & \textbf{13} (\scriptsize 0.09) \\
{$\text{LIB}[7]$} & 16 (\scriptsize 1.47) & timeout & \textbf{15} (\scriptsize 0.12) \\
{$\text{LIB}[8]$} & 18 (\scriptsize 5.95) & timeout & \textbf{17} (\scriptsize 0.20) \\
{$\text{LIB}[9]$} & 20 (\scriptsize 37.78) & timeout & \textbf{19} (\scriptsize 0.23) \\

\bottomrule
\end{tabular}
\caption{State and runtime (in seconds) comparison of our direct construction \texttt{GFM-GF} against our reduced automata, \texttt{Owl-Red} and \texttt{GFM-Red}. The runtime for \texttt{Owl-Red} and \texttt{GFM-Red} includes construction and consecutive application of our reduction (in seconds). Timeout is 300 seconds.}
\label{tab:evaluation-gfm-gf}
\end{table}

\section{Experiments for $\ltlfG \ltlfF \varphi$}
The first two LTL specification patterns above, namely \textbf{TDR} and \textbf{LIB}, refer to formulas that have the syntactic $\ltlfG \ltlfF \varphi$ structure and therefore, we can use them to evaluate our direct GFM automata construction.
In particular, we show below that the direct construction gives us a large improvement with respect to the standard state-of-the-art GFM constructions \cite{DBLP:conf/atva/KretinskyMS18, DBLP:conf/tacas/HahnPSS0W20}, even after applying our GFM reduction.

\paragraph{Experiment Setup.}
Table \ref{tab:evaluation-gfm-gf} compares our direct GFM construction (denoted by \texttt{GFM-GF}) to the automata obtained by our GFM state space reduction. Again, we denote number of states of reduced Owl and ``slim" GFM automata (\texttt{Owl-Red}, \texttt{Slim-Red}) but in contrast to Table \ref{tab:evaluation}, the runtime (in seconds), now contains both LTL to GFM construction and subsequent application of our reduction.

\paragraph{Results.}
For the LTL pattern TDR, Table \ref{tab:evaluation-gfm-gf} clearly shows that our direct construction \texttt{GFM-GF}, significantly outperforms both \texttt{Owl-Red} and \texttt{Slim-Red}, both in state size and runtime. 
Even when comparing the state size and construction time of \texttt{GFM-GF} to \texttt{Owl} and \texttt{Slim}, we can build exponentially more succinct automata, in a fraction of the time. 
For the LIB pattern, \texttt{GFM-GF} builds automata with a similar amount of states than \texttt{Owl-Red}, but again, we only require a fraction of the time. 
LIB[7] and LIB[8] timed out during reduction, while LIB[9] timed out during automata construction. More results can be found in Appendix~\ref{appendix:general-gfm-state-space}.

\section{Conclusion}

We conclude the paper by observing that 
our results immediately have an impact on a wide range of applications that use GFM automata, such as obtaining smaller strategies for planning problems, improving the efficiency of probabilistic verification~\cite{DBLP:conf/concur/HahnLST015,DBLP:conf/atva/SickertK16}, and reinforcement learning~\cite{DBLP:conf/tacas/HahnPSS0W20,deepltl}.
Furthermore, our results can be used to reduce the GFM automata obtained from variants of LTL, including LTLf+ and PPLTL+ recently proposed in \cite{ADRV2025,DLSW2Y2025}.

\section*{Acknowledgements}
We would like to thank the anonymous reviewers for their constructive feedback,
that helped improve the paper.
This work was supported in part by the CAS Project for Young Scientists in Basic Research (Grant No. YSBR-040), ISCAS Basic Research (Grant Nos. ISCAS-JCZD-202406, ISCAS-JCZD-202302), ISCAS New Cultivation Project ISCAS-PYFX-202201, the UKRI Erlangen AI Hub on Mathematical and Computational Foundations of AI (No. EP/Y028872/1) and by the EPSRC through grants EP/X03688X/1 and EP/X042596/1.

\bibliography{ref}

\clearpage

\appendix

\section{Related Work}

Note that our transformation procedure from a GFM automaton $\Agfm$ to a 0/1-PA $\Apa$ is similar to the one that converts a unambiguous \buchi automaton (UBA) to a 0/1-PA proposed in~\cite{LiPST25}.
Nonetheless, \cite{LiPST25} considers only the UBAs as input while we consider GFM automata.
Further, \cite{LiPST25} works only on Markov chains while we consider MDPs, which is a more general model and thus makes our algorithm also work on MCs in probabilistic verification.
Hence, our work can be seen as a generalisation to~\cite{LiPST25} in this regard.

According to~\cite[Theorem 10]{DBLP:conf/concur/Schewe0Z23}, minimising GFM automata is PSPACE-hard.
Instead of proposing GFM minimisation algorithms, we propose to use polynomial GFG minimisation to reduce the input GFM automata to 0/1-PAs, which although do not accept the original languages but can be used for analysing MDPs with our product MDP definition.

\cite{DBLP:conf/lata/KleinMBK14} showed that GFG automata can be used for analysing MDPs but current constructions give GFG automata even larger than their deterministic counterparts.
This hinders the use of GFG automata for probabilistic verification in practice.
Our work is the first successful effort to combine the strengths of GFG automata and GFM automata.
That is, we use the polynomial GFG minimisation algorithm~\cite{AbuRadiK22} to obtain automata that are good for MDPs.
Our algorithm not only gives smaller strategies for stochastic MDP planing problems but also have the potential to be more efficient in particular when the input MDPs are large.

\section{Action Indexing}
\label{appendix:action-indexing}
We note that the action set $\act\times \states$ is often used for $\M\times \A$ in other literature, e.g.~\cite{DBLP:conf/concur/Schewe0Z23}.
That is, the traditional product $\M\times \A$ resolves the nondeterminism in $\A$ by asking the agent to make the choice of the successors $q$ in the action $\langle \action, q\rangle$.
Our observation is that, instead of explicitly storing the \emph{selected} successor in the action names, we only need to know the relative position $i$ of the chosen successor in the predefined order of $\trans(q, \letter)$, where $q \in \states$ and $\letter \in \alphabet$.
To uniquely associate a successor $q'$ with its relative position $i$ in $\trans(q, \letter)$, we need an order over the successors.
A simple way to define an order is to use a \emph{global} order over $\states$, such as ordering $\states = \{q_0, \cdots, q_n\}$ by their index names, but we can also define a different \emph{local} order for each pair of $q$ and $\letter$.
We assume that for a state $q$ and a letter $\letter$ in $\A$, we have a function $\order_{\A, q, \letter}: \trans(q, \letter) \rightarrow \mathbb{N}$ that maps a state $q'$ in $\trans(q, \letter)$ of $\A$ to its unique position $i \in [|\trans(q, \letter)|]$, i.e. $\order_{\A, q,\letter}(q') = i$.
In this way, our product definition is in fact equivalent to the one using the action set $\act\times\states$, but has a slight advantage of using fewer actions.

\section{Non-GFM Example}
\label{app:non-gfm}

\begin{figure}[t]
    \centering
      \begin{tikzpicture}[
          on grid, auto,
          node distance=15mm, 
          draw=black,
          every initial by arrow/.style={thick},
          >={Stealth}, 
          state/.style={circle, draw, thick, minimum size=7mm} 
        ]
          \node[state,initial,initial text={}] (q0) {$q_0$};
        
          \node[state,above right=10mm and 25mm of q0] (q1) {$q_1$};
          
          \node[state,below right=10mm and 25mm of q0] (q2) {$q_2$};
        
          \path[->,thick]
            (q0) edge[bend left=20]  node[above, sloped] {$a,b,c$} (q1)
            (q0) edge[bend right=20] node[below, sloped] {$a,b,c$} (q2)
            (q1) edge[loop above, double] node {$b$} (q1)
            (q2) edge[loop below, double] node {$c$} (q2);
        \end{tikzpicture}
    \caption{NBA example not good-for-MDP}
    \label{fig:non-gfm-1}
\end{figure}
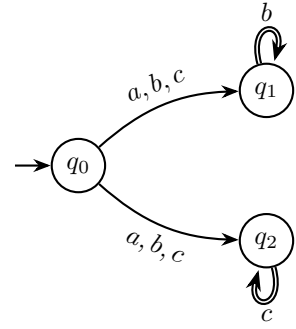

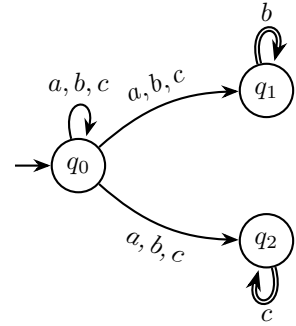
\begin{figure}
    \centering
          \begin{tikzpicture}[
          on grid, auto,
          node distance=15mm, 
          draw=black,
          every initial by arrow/.style={thick},
          >={Stealth}, 
          state/.style={circle, draw, thick, minimum size=7mm} 
        ]
          \node[state,initial,initial text={}] (q0) {$q_0$};
        
          \node[state,above right=10mm and 25mm of q0] (q1) {$q_1$};
          
          \node[state,below right=10mm and 25mm of q0] (q2) {$q_2$};
        
          \path[->,thick]
            (q0) edge[bend left=20]  node[above, sloped] {$a,b,c$} (q1)
            (q0) edge[bend right=20] node[below, sloped] {$a,b,c$} (q2)
            (q1) edge[loop above, double] node {$b$} (q1)
            (q2) edge[loop below, double] node {$c$} (q2)
            (q0) edge[loop above] node {$a,b,c$} (q0);
        \end{tikzpicture}
    \caption{NBA example good-for-MDP}
    \label{fig:non-gfm-2}
  \end{figure}

Figure \ref{fig:non-gfm-1} and Figure \ref{fig:non-gfm-2} display two nondeterministic \buchi automata. The NBA in Figure \ref{fig:non-gfm-1} is not GFM, while the NBA in Figure \ref{fig:non-gfm-2} is GFM.
The NBA of Figure \ref{fig:non-gfm-1} is not GFM since once the agent picks $q_1$, the adversary can force only $c$ loops. Note instead, adding to the NBA of Figure \ref{fig:non-gfm-1} a self loop with all letters on $q_0$ makes it GFM (see Figure \ref{fig:non-gfm-2}), though it then recognises a different language.

\section{Proof of Theorem~\ref{thm:pa-gfm}}
\label{app:pa-gfm}

\begin{proof}

    Obviously, by definition, $\psem(\M, \mathcal{D}) = \psem(\M, \pa)$ since $\lang{\pa} = \lang{\mathcal{D}}$.

    Since $\D$ is deterministic and thus GFM, $\psyn(\M, \mathcal{D}) = \psem(\M, \mathcal{D})$.
    Therefore, we only need to show that $\psyn(\M, \pa) = \psem(\M, \mathcal{D})$.

    First, we prove that $\psem(\M, \mathcal{D}) \leq \psyn(\M, \pa)$.
    Let $\strategy$ be the optimal strategy for $\M$ to achieve $\psem(\M, \mathcal{D})$.
    We then can apply $\strategy$ to resolve the nondeterminism in the actions and obtain a MC $\M^{\strategy}$.
    Further, $\psyn(\M^{\strategy}, \D) = \psem(\M^{\strategy}, \D) = \psem(\M, \D)$ since $\strategy$ is an optimal strategy.
    This in fact gives a product $\M^{\strategy} \otimes \pa$ as the strategy $\strategy$ does not rely on the 0/1-PA $\pa$.
    According to \cite[Proposition~2]{LiPST25}, the product $\M^{\strategy} \otimes \pa$ achieves the maximal semantic probability $\psem(\M^{\strategy}, \pa)$, i.e., $\psyn(\M^{\strategy}, \pa) = \psem(\M^{\strategy}, \pa)$.
    This entails that $\psem(\M^{\strategy}, \pa) \leq \psyn(\M, \pa)$.
    That is, we have $\psem(\M^{\strategy}, \mathcal{D}) = \psem(\M^{\strategy}, \pa) \leq \psyn(\M, \pa)$ since $\lang{\mathcal{D}} = \lang{\pa}$.
    Again, since $\strategy$ is the optimal strategy and $\M^{\strategy}$ is the induced MC, we have $\psem(\M^{\strategy}, \mathcal{D}) = \psem(\M, \mathcal{D})$.
    It then follows that $\psem(\M, \mathcal{D}) \leq \psyn(\M, \pa)$.

    We now show that $\psyn(\M, \pa) \leq \psem(\M, \D)$. 
    Let $\strategy$ be any optimal finite memory strategy on $\M$ such that $\psyn(\M^{\strategy}, \pa) = \psyn(\M, \pa)$.
    According to \cite[Proposition~2]{LiPST25}, $\pa$ can preserve the semantic satisfaction probability for MCs, i.e., we then have $\psyn(\M^{\strategy}, \pa) = \psem(\M^{\strategy}, \pa) $ since $\M^{\strategy}$ is an MC.
    Note that our product MDP definition de-generalises to the product MC definition in~\cite{LiPST25} given a strategy $\strategy$ on $\M$.
    Together with the fact that $\lang{\pa} = \lang{\D}$, it holds that $\psyn(\M^{\strategy}, \pa) = \psem(\M^{\mu}, \D)$ since $\psem(\M^{\strategy}, \D) = \psem(\M^{\strategy}, \pa)$.
    Thus, $\psyn(\M^{\strategy},\pa) = \psyn(\M^{\strategy}, \D) \leq \psem(\M, \D)$ as $\strategy$ might not be the optimal strategy for $\M$ to achieve $\psem(\M, \D)$. 
    It then immediately follows that $\psyn(\M, \pa) \leq \psem(\M, \D)$ since for all $\strategy$, we have $\psyn(\M^{\strategy}, \pa) \leq \psem(\M, \D)$.

    Therefore, $\psyn(\M, \pa) = \psem(\M, \mathcal{D})$.
    We have thus completed the proof. 
\end{proof}

\paragraph{Equivalent 0/1-PA for Every DBA}
Notice, that every DBA can be transformed into an equivalent 0/1-PA by simply transitioning to the only successor with probability one.

\section{Proof of Lemma~\ref{lem:syn-same-prob}}

Observe that $\Adba$ only make the nondeterminism of $\Agfm$ explicit.
While in the classical definition of $\M\ \times \Agfm$, we can see that the action set $\act$ is extended to $\act \times [k]$ so to resolve the nondeterminism in $\A$ by selecting different successor via actions.
This is also in fact a way of resolving nondeterminism in $\A$.

In our constructed DBA $\Adba$, for each state $q \in \states_{\DBA}$ and a letter $\letter \in \alphabet$, $q \in \trans_{\text{GFM}}(q, \letter)$ if, and only if, there exists a unique integer $i \in [k]$ such that $q' = \trans_{\text{DBA}}(q, \langle \letter, i\rangle )$.
It is easy to see that we have 
$\prob^{\times}_{\text{GFM}}(\langle s, q\rangle, \langle \action, i\rangle, \langle s', q'\rangle) > 0$ in $\M\times \A$ if, and only if, we have $\prob^{\times}_{\text{DBA}}(\langle s, q\rangle, \langle \action, i, 1\rangle, \langle s', q'\rangle) > 0$ in $\M'\times \Adba$ for all $s, s', q, \action$, where $q' = \delta_{\DBA}(q, \langle \lab(s, \action), i\rangle)$.
Further, there is one-to-one correspondence between the two transitions $(\langle s, q\rangle, \langle \action, i\rangle, \langle s', i\rangle)$ of $\M\times \A$ and $(\langle s, q\rangle, \langle \action, i, 1\rangle, \langle s', q'\rangle)$ of $\M'\times \Adba$, with $\prob^{\times}_{\text{GFM}}(\langle s, q\rangle, \langle \action, i\rangle, \langle s', q'\rangle) = \prob^{\times}_{\text{DBA}}(\langle s, q\rangle, \langle \action, i, 1\rangle, \langle s', q'\rangle) = \prob(s, \action, s')$.
Since $\Agfm$ is GFM, there is a memoryless strategy from every state in $\M \times \Agfm$ to obtain the corresponding optimal satisfaction probability. 
Therefore, from a memoryless optimal strategy on $\M\times\Agfm$, we can immediately construct an optimal strategy on $\M' \times \Adba$.
On the other hand, the product $\M'\times \Adba$ is essentially the same as $\M \times \Agfm$. 
Hence, from every strategy $\strategy'$ of a state $\langle s, q\rangle $ in $\M' \times \Adba$, regardless of memoryless or not, we can naturally construct a strategy $\strategy $ of the state $\langle s, q\rangle $ in $\M\times \Agfm$ obtaining the same satisfaction probability.

Therefore, the lemma holds.

\paragraph{Optimal values coincide.}
By the definition of GFM automata, the optimal strategy on $\mathcal M \times \mathcal A_{\mathrm{GFM}}$ equals the optimal satisfaction probability of $\varphi$ in $\mathcal M$.
As shown in Lemma ~\ref{lem:syn-same-prob}, there is a one-to-one correspondence between the two product MDPs $\M \times \Agfm$ and $\M'\times \Adba$, with essentially identical state spaces, transitions, and probabilities. Hence, both yield the same optimal satisfaction probability.

\section{Proof of Lemma~\ref{lem:pa-lang}}

Lemma~\ref{lem:pa-lang} directly follows from~\cite[Lemma 3]{LiPST25}.
In order to be self-contained, we briefly sketch the proof idea here.
From a DBA $\Adba$ and thus a DCA $\Adca$, we obtain a minimal GFG NCA $\Agfgmin$ whose language is exactly $\lang{\Adca}$.
$\Agfgmin$ are both
semantically deterministic and safe deterministic.
As mentioned earlier in the paper, $\Agfgmin$ is semantically deterministic, if for any state $q $ and letter $\letter$, all successors have the same language, i.e., for two states $s, t \in \trans_{\text{GFGMin}}(q, \letter)$, we have $\lang{\Agfgmin^s} = \lang{\Agfgmin^t}$.
$\Agfgmin$ is also safe deterministic if when removing all $\acccond_{\text{GFGMin}}$-transitions, all strongly connected components are deterministic.
Since $\Agfgmin$ has a co-\buchi condition, every accepting run will eventually enter the so-called deterministic safe component because $\Agfgmin$ is safe-deterministic.

For any word $w \in \lang{\Adba}$, which is of course not in $\lang{\Adca}$ (and thus not in $\lang{\Agfgmin}$) as $\Adba$ and $\Adca$ complement each other, all runs in $\Agfgmin$ over $w$ are not accepting.
Hence, when $\Agfgmin$ is treated as an NBA $\Anba$, all runs over $w$ are accepting.
Therefore, the probability measure of the accepting runs over an accepting word $w$ in $\Apa$ is one.
Now, we show that the probability measure of accepting runs over a rejecting word $w \notin \lang{\Adba}$ is zero.
This is also easy as $\Agfgmin$ is semantically deterministic.
Let $w \notin \lang{\Adba}$ and thus we have $w \in \lang{\Agfgmin}$.
There must be an accepting run $q_0 q_1 q_2 \cdots$ over $w$ in $\Agfgmin$.
For any run $q'_0 q'_1 q'_2 \cdots$, we know that $\lang{\Agfgmin^{q'_i}} = \lang{\Agfgmin^{q_i}}$ for all $i \geq 0$ since $\Agfgmin$ is semantically deterministic.
Therefore, every state $q'_i$ has a finite path to a deterministic safe component since from $q_i$, the remaining suffix of $w$ will be accepted.
Hence, it is always possible for a nonaccepting run in $\Agfgmin$ to be changed to an accepting run at any time for a positive probability.
That is, every run in $\Anba$ over $w \notin \lang{\Adba}$ has a positive probability to reach a safe deterministic component at any time.
So, the probability measure of accepting runs over $w \notin \lang{\Adba}$ must be zero since it is always with positive probability to escape to nonaccepting zone in $\Anba$.

It then follows that the $\Apa$ is 0/1-PA and recognises the language $\lang{\Adba}$.

\section{Preservation of $\omega$-Regular Properties under GFG minimisation}
\label{app:omega-reg-preservation}
GFM automata recognise all $\omega$-regular languages~\cite{DBLP:conf/tacas/HahnPSS0W20}. In our reduction pipeline, we first transform these GFM automata into DBAs by making the nondeterministic choices explicit, using additional letters and then transform them into 0/1-PAs. Note that the original $\omega$-regular language can always be recovered by simply ignoring these extra letters. Hence, the $\omega$-regular properties remain fully preserved throughout our reduction.

\section{Proof of Theorem~\ref{thm:pa-main}}
\label{app:pa-main}
\begin{proof}
    As a trivial result of Theorem~\ref{thm:end-component}, within an accepting MEC, there is \emph{memoryless} and random strategy to visit all state-actions pairs with probability one. 
Therefore, we can treat all accepting MECs as sink goal states.
Since there exists a \emph{memoryless} optimal strategy for reachability goals, we have a memoryless strategy on $\M'\otimes \pa$ to solve the stochastic planning problem.
From the optimal strategy $\strategy_{\M'}$ for $\M' \otimes \pa$, we can obtain the optimal strategy for $\M'$ to achieve $\psem(\M', \pa)$.
Further, from an optimal strategy $\strategy_{\M'}: S\times \states_{\pa} \rightarrow \distr(\act')$, we can obtain an optimal strategy $\strategy_{\M}: S\times \states_{\pa} \rightarrow \distr(\act)$ by setting $\strategy_{\M}(\langle s, q\rangle)(\action) = \Sigma_{i \in [k]}\strategy_{\M'}(\langle s, q\rangle)(\langle \action, i\rangle)$ to achieve $\psem(\M', \Adba) = \psem(\M', \pa)$.
Further, by Lemma~\ref{lem:syn-same-prob}, we know that a strategy that maximises the satisfaction probability of $\lang{\Adba}$ from the initial state in $\M'$ is also a strategy that maximises the satisfaction probability of $\lang{\A}$ in the same initial state in $\M$.
Since $\psem(\M, \A) = \psem(\M', \Adba)$, we then have obtained the optimal strategy by letting $\strategy' = \strategy_{\M}$. 
\end{proof}

\section{Proof Lemma~\ref{reapeated-reachability}}
\begin{proof}
\label{app:proof-reapeated-reachability}
    It is easy to see that the right-hand side implies the left-hand side.
    If $P$ is specifiable in $\ltlfG\ltlfF \varphi $ where $\varphi$ is a co-safety formula, then for any $w \in P$, we have that $w_i \models \varphi$ for infinitely many integers $i > 0$.
    Let $w[i \cdots i'] \models \varphi$ for infinitely many $i$ and $i'$ with $i \leq i'$.
    We can let $ [\varphi] = R \cdot \ttrue^{\omega}$.
    Then, we know that for any $w \in P$, there are infinitely many prefixes of $w$ belong to $\finwords\cdot R$.
    Therefore, $P$ is a repeated reachability property specifiable in LTL.
    
    The implication from the left to the right is an immediate result from~\cite[Theorem 3]{DBLP:conf/lics/SickertE20}.
    Note that the strong release temporal operator $\textsf{M}$ can be rewritten as $\varphi~\textsf{M}~\psi \equiv \psi~\ltlfU~(\varphi\land \psi)$.
    So, any co-safety/reachability/guarantee property in $R$ can be specified by a co-safety formula $\varphi$.
    To make it repeat infinitely often as prefixes, we can just add $\ltlfG\ltlfF$ in front of $\varphi$.
\end{proof}

\section{Proof of Theorem~\ref{thm:GF-fragment-construction}}

\begin{proof}

Let $\varphi$ be a $\ltlfG$-free formula, i.e., a formula that does not contain $\ltlfG$ modality and $w$ be a word.
Then $w \models \varphi$ if, and only if, there exists an integer $j > 0$, $\af(\varphi, w[0\cdots j]) = \ttrue$, 
according to \cite[Lemma 11]{DBLP:conf/cav/EsparzaK14}. The function $\af$ is defined as in \cite{DBLP:conf/cav/EsparzaK14}.

Recall that for a co-safety property formula $\varphi$, one can construct an NFA $\N_{\varphi} = (\states, q_0, \trans, F)$ such that $\finlang{\N_{\varphi}}\cdot (\ttrue)^{\omega} = [\varphi]$ where $\finlang{\N}$ denotes the set of finite traces accepted by ${\N_{\varphi}}$.
Note that the alphabet here is $\alphabet = 2^{\ap}$.

Recall now the construction for the GFM automaton $\A$ from $\N_{\varphi}$.
We construct from $\N_{\varphi}$ a GFM automaton $\A = ( \states_{\A}, q_{\A}, \trans_{\A}, \acccond_{\A})$ for $\ltlfG\ltlfF \varphi$ where 
\begin{itemize}
\item $\states_{\A} \subseteq \left((\states \setminus F) \cup \{q_0\}\right)$, $q_{\A} = q_0$, 
\item $\trans_{\A}: \states_{\A} \times \alphabet \rightarrow 2^{\states_{\A}}$ is defined such that for each $q \in \states_{\A}$ and $\letter \in \alphabet$, we have 
(1) $q' \in \trans_{\A}(q,\letter)$ if $q' \in \trans(q, \letter)$ with $q' \notin F$, and (2) $q_0 \in \trans_{\A}(q, \letter)$, and
\item 
 $\acccond_{\A} = \{(q,\letter,q_0) \in \Delta(\trans_{\A}) \mid \exists q' \in \trans(q, \letter) \land q' \in F\}$.
\end{itemize}

We first prove that $\lang{\A} = [\ltlfG\ltlfF \varphi]$.
Assume that $w \models \ltlfG\ltlfF \varphi$.
By definition, for any given integer $i > 0$, there exists a position $j \geq i$ such that $w_j \models \varphi$.
Assume that $\mathbf{J} = \{j > 0 \mid w_j \models \varphi\} = \{j_0, j_1, \cdots \}$.
We can construct a strategy to produce an accepting run $\rho$ for $w$ in $\A$ as follows:
before reaching position $j_0$, we choose to stay in the initial state $q_0$.
Since $w_{j_0} \models \varphi$, there must exist a position $j'_0 \geq j_0$ such that $\af(\varphi, w[j_0\cdots j'_0]) = \ttrue$.
So, we know that $\delta(q_0, w[j_0\cdots j'_0]) \cap F \neq \emptyset$, i.e., there must be an accepting finite run to a final state, say $f_0$.
By construction, we are able to transition back to $q_0$ when we read the word $w[j_0\cdots j'_0]$ instead of reaching the final state $f_0$.
The corresponding transition back to $q_0$ is accepting and in $\acccond_{\A}$.
Next, we select the smallest integer $j_k \in \mathbf{J}$ such that $j_k > j'_0$.
We choose to loop on the initial state $q_0$ until we reach position $j_k$ and then move according to the $\trans$ function afterwards.
Analogously, we are able to go back to the initial state $q_0$ via an accepting transition because $w_{j_k} \models \varphi$.
If we repeat this strategy, in the end, the constructed run $\rho$ will visit accepting transitions for infinitely often.
Therefore, $w  \in \lang{\A}$.
That is, we have $[\ltlfG\ltlfF \varphi] \subseteq \lang{\A}$.

The proof for the other direction that $\lang{\A} \subseteq [\ltlfG\ltlfF \varphi]$ is straightforward.
This is because as soon as the accepting transition is visited, the finite word from the latest initial state to the accepting transition must satisfy $\varphi$.
Therefore, if an accepting run visits accepting transitions for infinitely many times, the corresponding word also satisfies $\varphi$ at infinitely many positions.

It then follows that $\lang{\A} = [\ltlfG\ltlfF \varphi]$.

Now we have to prove that $\A$ is GFM.
We can actually prove a stronger result that $\A$ is stochastically resolvable automaton in a sense that we can resolve the nondeterminism randomly and obtain a language equivalent 0/1-PA $\pa$.
This follows almost directly from our construction.
 
We prove that $\pp_{\pa}(w) = 1$ for each word $w \in \lang{\pa}$ and $\pp_{\pa}(w) = 0$ for $w \notin \lang{\A}$.
Let $w \in \lang{\A}$.
The whole transition graph of $\A$/$\pa$ is a strongly connected component.
Therefore, every finite path will be visited almost surely because the nondeterminism is resolved randomly.
It then follows that all transitions will be visited infinitely often.
This then entails that a run of $\pa$ over $w$ almost surely visits an accepting transition.
Hence, $\pp_{\pa}(w) = 1$.
Let $w \notin \lang{\A}$.
By semantics, we know that there exists an integer $k > 0$ such that for all $j \geq k$, we have that $w_j \models \neg \varphi$.
This means that every run of $\pa$ over $w$ must not visit the accepting transition from some point on, which would only happen with probability zero.
Therefore, we have $\pp_{\pa}(w) = 0$ for a word $w \notin \lang{\A}$.

It then follows that $\pa$ is a language-equivalent 0/1-PA of $\A$.
Therefore, $\A$ is GFM according to our previous algorithm.

\qedhere

\end{proof}

\section{Need for Co-Safety in $\ltlfG \ltlfF \varphi$}
\label{app:co-safety-need}
For repeated reachability formulas $\ltlfG \ltlfF \varphi$, $\varphi$ are co-safety formulas that go beyond the intersection of safety and co-safety e.g. $(a \ltlfU b)$~\cite{DBLP:journals/jacm/EsparzaKS20}. The key property we exploit of these formulas is that they can be recognised by a reachability automaton (NFA in our proof). Any specification that has this characteristic can be handled by our method.

\section{Optimal Strategy and Further Optimisations}
\label{appendix:optimal-strategy-further-optimisations}
We give the following 
Algorithm~\ref{alg} to synthesise the optimal strategy to achieve $\psem(\M', \Adba)$ in more details. 

\begin{algorithm}[t]
\caption{Synthesising an Optimal Strategy for an MDP $\M$ against an \ltl specification $\varphi$}\label{alg}
\begin{algorithmic}[1]

\STATE Construct a GFM automaton $\A$ from $\varphi$.
\STATE Create $\M'$ from $\M$ and  $\Adba$ (and thus $\Adca$) from $\A$.
\STATE Apply GFG minimisation on $\Adca$ and treat the minimised automaton $\Agfgmin$ as a \buchi automaton $\Anba$.

\STATE Resolve the nondeterministic choices in $\Anba$ with random choices, yielding a 0/1-PA $\Apa$.
\item Create $\M' \otimes \Apa$.
\STATE Identify the MECs with accepting transitions in the product MDP and mark the states in accepting MECs as accepting states, i.e., reducing it to reachability goals.
\STATE\label{step:reach-goals} Determine the chance of reaching an accepting state in $\M' \otimes \Apa$, recording a strategy for those states that reach it with a probability in $(0,1]$, i.e., solving reachability planning problems.
\STATE For the states in the accepting MECs, use the strategy to maximally randomise among all successors that stay within the accepting region, i.e., using a memoryless strategy to visit all transitions in the accepting MECs.

\end{algorithmic}
\end{algorithm}
We can optimise the quantitative analysis after finding the accepting MECs (Step~3 of Algorithm~\ref{alg}) by exploiting that, for the 0/1-PA $\Apa$, the update of the right language of states is deterministic.
(Recall that $\Apa$ is semantically deterministic.)
The right language is the language accepted from that state in $\Apa$.
We call the deterministic update automata of these languages $\R$ and note that $\R$ can be viewed as a quotient automaton of $\Apa$.

We can simply make every state $(s, r)$ in $\M'\times \R$ accepting if there is a state $(s, p)$ in an accepting MEC of $\M'\otimes \Apa$ such that $p$ and $r$ are language equivalent in $\Apa$, and then calculating the chance of winning for the remaining states by maximising the chance of reaching any such state $(s, r)$.

This provides an optimal strategy which operates on $\M' \times \R$ until such a state $(s,r)$ is reached, and then moves to a random strategy on $\M' \otimes \Apa$ on such an accepting MEC.

\section{Automata Completion and Sink States}
\label{app:at-completion}
We deliberately refrained from completing all automata returned by \texttt{Owl}. This ensures that our setup remains fully reproducible and consistent with how the original tools provide their outputs.
Completing them would only add one single (sink) state, without adding meaningful insight. Therefore, we keep the original automaton to maintain transparency and reproducibility.

\section{Additional Experiments}
\label{appendix:general-gfm-state-space}
Tables \ref{tab:appendix-table-1} and \ref{tab:appendix-table-2} show the effectiveness of our GFM state-space reduction on additional LTL formulas, extracted from Literature. 
Table \ref{tab:appendix-table-3} demonstrates the applicability of our direct construction for the $\ltlfG\ltlfF\varphi$ fragment.
Where applicable, we use \texttt{GENLTL} of Spot \cite{DBLP:conf/atva/Duret-Lutz13} and existing LTL datasets from the Owl \cite{DBLP:conf/atva/KretinskyMS18} repository.

We use a sideway layout to present each LTL formula in full, together with the number of automata states. Using the same notation as in the main-body, we denote Owl \cite{DBLP:conf/atva/KretinskyMS18} by \texttt{Owl}, and the ``slim" GFM construction of \cite{DBLP:conf/tacas/HahnPSS0W20} by \texttt{Slim}. \texttt{Owl-Red} and \texttt{Slim-Red} denote the subsequent application of our reduction. \texttt{GFM-GF} denotes our direct construction. We measure runtime in seconds and report a timeout after 300 seconds. Our reduction returns complete automata with an explicit sink state, which can add one extra state.

Furthermore, we introduce following novel pattern:
\\
\noindent
{\textbf{Nested Conjunction Sequence (NCS).}} 
Conceptually inspired by \cite{DBLP:journals/corr/abs-1709-02102}, we define
$\text{NCS}[k_1, \dots, k_n] = \ltlfG\ltlfF(a \land \ltlfX^{k_1}b \land \ltlfX^{k_1 + k_2}c \land \ltlfX^{k_1 + k_2 +k_3} d \land \ltlfX^{k_1 + k_2 +k_3+k_4} e \land \dots)$, where $\ltlfX^n$ denotes $n$ nested applications of the next operator.


\begin{sidewaystable}[]
\vspace{9.5cm}
\small
\setlength{\tabcolsep}{4pt}
\renewcommand{\arraystretch}{1.1}
\begin{tabular}{
    >{\raggedright\arraybackslash}p{4.5cm}    
    >{\raggedright\arraybackslash}p{10.5cm}   
    >{\centering\arraybackslash}p{1.75cm}     
    >{\centering\arraybackslash}p{1.75cm}     
    >{\centering\arraybackslash}p{1.75cm}     
    >{\centering\arraybackslash}p{1.75cm}     
}
\toprule
  Reference
& Formula
& \texttt{Owl}
& \texttt{Owl-Red}
& \texttt{Slim}
& \texttt{Slim-Red}\\
\midrule

\multirow{2}{*}{\cite{DBLP:conf/fmsp/DwyerAC98}} & 
\tiny\texttt{G(!a|G!b|((!c|X(!bU(d\&Fe))|X(bR!d))Ub))} & 23 (\scriptsize 0.46) & 17 (\scriptsize 0.01) & 21 (\scriptsize 0.09) & \textbf{16} (\scriptsize 0.01) \\ &
\tiny\texttt{G(!a|((!b|X(!cU(d\&Fe))|X(cR!d))U(c|G(!b|X(!cU(d\&Fe))|X(cR!d)))))} & 39 (\scriptsize 0.47) & 23 (\scriptsize 0.03) & 20 (\scriptsize 0.09) & \textbf{15} (\scriptsize 0.01) \\

\midrule

\multirow{1}{*}{\cite{DBLP:conf/concur/EtessamiH00}} & 
\tiny\texttt{aU(b\&X(c\&F(d\&XF(e\&XF(f\&XFg)))))} & 13 (\scriptsize 0.43) & \textbf{9} (\scriptsize 0.01) & 127 (\scriptsize 0.14) & 54 (\scriptsize 0.31) \\

\midrule

\multirow{2}{*}{\cite{DBLP:journals/fmsd/TabakovRV12}} &
\tiny\texttt{G(p1->(p1U(p2\&(p2U(p3\&(p3U(p4\&(p4Up5))))))))} & 44 (\scriptsize 1.16) & \textbf{23} (\scriptsize 0.02) & 51 (\scriptsize 0.22) & 39 (\scriptsize 0.02) \\ &
\tiny\texttt{G(p1->(p1U(p2\&(p2U(p3\&(p3U(p4\&(p4U(p5\&(p5Up6))))))))))} & 150 (\scriptsize 1.41) & \textbf{56} (\scriptsize 0.17) & 232 (\scriptsize 0.82) & 159 (\scriptsize 0.76) \\ &
\tiny\texttt{G(p1->(p1U(p2\&(p2U(p3\&(p3U(p4\&(p4U(p5\&(p5U(p6\&(p6Up7))))))))))))} & 433 (\scriptsize 5.92) & \textbf{249} (\scriptsize 27.75) & 1425 (\scriptsize 12.24) & 753 (\scriptsize 163.32) \\

\midrule

\multirow{10}{*}{\cite{DBLP:conf/atva/Duret-Lutz13}} &
\tiny\texttt{G(!a|(aU(b\&(bU(c\&(cUd))))))} & 14 (\scriptsize 0.48) & \textbf{10} (\scriptsize 0.01) & 12 (\scriptsize 0.09) & 13 (\scriptsize 0.01) \\ &
\tiny\texttt{(FGa|GFb)\&(FGc|GFa)\&(FGd|GFc)} & 14 (\scriptsize 0.43) & \textbf{13} (\scriptsize 0.01) & 78 (\scriptsize 0.11) & 69 (\scriptsize 0.06) \\ &
\tiny\texttt{(FGa|GFb)\&(FGc|GFa)\&(FGd|GFc)\&(FGe|GFd)} & 30 (\scriptsize 0.44) & \textbf{27} (\scriptsize 0.01) & 406 (\scriptsize 0.51) & 370 (\scriptsize 5.54) \\ &
\tiny\texttt{GF((a\&XXXa)|(!a\&XXX!a))} & 17 (\scriptsize 0.43) & 18 (\scriptsize 0.01) & 17 (\scriptsize 0.09) & \textbf{10} (\scriptsize 0.01) \\ &
\tiny\texttt{GF((a\&XXXXa)|(!a\&XXXX!a))} & 33 (\scriptsize 0.44) & 34 (\scriptsize 0.01) & 33 (\scriptsize 0.09) & \textbf{18} (\scriptsize 0.01) \\
& \tiny\texttt{a\&X(b|c)\&(!bU(b\&X(c\&XXXXGa)))\&G(!c|X((d|e)\&Xf))\&G(!f|X((d|e)\&X(a\&X(b|c|Ga))))\&F(\newline a\&X(!a\&(((!c|X((d\&F(b\&F(c\&Xd)))|(e\&F(b\&F(c\&Xe)))))\&(!f|X((d\&F(b\&F(f\&Xd)))|(e\&\newline F(b\&F(f\&Xe))))))Ua)))\&G((!c|!d)\&(!a|!b)\&(!a|!c)\&(!a|!d)\&(!b|!c)\&(!b|!d)\&(!d|!e)\&\newline (!a|!e)\&(!b|!e)\&(!c|!e)\&(!d|!f)\&(!e|!f)\&(!a|!f)\&(!b|!f)\&(!c|!f))} & 163 (\scriptsize 140.13) & \textbf{148} (\scriptsize 0.02) & 149 (\scriptsize 0.17) & \textbf{148} (\scriptsize 0.01) \\ &
\tiny\texttt{G((!c|!d)\&(!a|!b)\&(!a|!c)\&(!a|!d)\&(!b|!c)\&(!b|!d))\&(!aU(a\&X((b|c)\&XGd)))\&F(d\&XXd\newline \&((Xb\&G(!a|Xb))|(Xc\&G(!a|Xc))))} & 19 (\scriptsize 0.60) & 17 (\scriptsize 0.01) & 20 (\scriptsize 0.09) & \textbf{16} (\scriptsize 0.01) \\ &
\tiny\texttt{G((!c|!d)\&(!a|!b)\&(!a|!c)\&(!a|!d)\&(!b|!c)\&(!b|!d))\&(!bU(b\&X((c|d)\&X((c|d)\&XGa))))\newline \&F(a\&((Xc\&G(!b|Xc))|(Xd\&G(!b|Xd)))\&XXXa\&((XXc\&G(!b|XXc))|(XXd\&G(!b|XXd))))} & 117 (\scriptsize 4.84) & 108 (\scriptsize 0.02) & 147 (\scriptsize 0.11) & \textbf{107} (\scriptsize 0.02) \\

\midrule

\multirow{6}{*}{\cite{DBLP:conf/spin/GeldenhuysH06}} & 
\tiny\texttt{(Fp1|Gp2)\&(Fp2|Gp3)\&(Fp3|Gp4)\&(Fp4|Gp5)\&(Fp5|Gp6)\&(Fp6|Gp7)\&(Fp7|Gp8)} & --  & --  & 1690 (\scriptsize 3.17) & \textbf{612} (\scriptsize 204.61) \\ &
\tiny\texttt{(GFp1|FGp2)\&(GFp2|FGp3)\&(GFp3|FGp4)\&(GFp4|FGp5)} & 30 (\scriptsize 0.46) & \textbf{27} (\scriptsize 0.02) & 406 (\scriptsize 0.54) & 370 (\scriptsize 5.70) \\ &
\tiny\texttt{(GFp1|FGp2)\&(GFp2|FGp3)\&(GFp3|FGp4)\&(GFp4|FGp5)\&(GFp5|FGp6)} & 66 (\scriptsize 0.49) & \textbf{58} (\scriptsize 0.03) & 3074 (\scriptsize 9.09) & timeout \\ &
\tiny\texttt{(GFp1|FGp2)\&(GFp2|FGp3)\&(GFp3|FGp4)\&(GFp4|FGp5)\&(GFp5|FGp6)\&(GFp6|FGp7)} & 146 (\scriptsize 1.01) & \textbf{127} (\scriptsize 0.14) & timeout  & --  \\ &
\tiny\texttt{(GFp1|FGp2)\&(GFp2|FGp3)\&(GFp3|FGp4)\&(GFp4|FGp5)\&(GFp5|FGp6)\&(GFp6|FGp7)\&(GFp7|FGp8)} & 322 (\scriptsize 3.78) & \textbf{279} (\scriptsize 0.96) & timeout & --  \\ &
\tiny\texttt{(GFp1|FGp2)\&(GFp2|FGp3)\&(GFp3|FGp4)\&(GFp4|FGp5)\&(GFp5|FGp6)\&(GFp6|FGp7)\&(GFp7|FGp8)\newline \&(GFp8|FGp9)} & 706 (\scriptsize 23.72) & \textbf{612} (\scriptsize 8.78) & timeout & --  \\ &
\tiny\texttt{(GFp1|FGp2)\&(GFp2|FGp3)\&(GFp3|FGp4)\&(GFp4|FGp5)\&(GFp5|FGp6)\&(GFp6|FGp7)\&(GFp7|FGp8)\newline \&(GFp8|FGp9)\&(GFp9|FGp10)} & 1538 (\scriptsize 226.73) & \textbf{1337} (\scriptsize 76.20) & timeout & -- \\

\midrule

\multirow{2}{*}{\cite{DBLP:journals/corr/abs-1709-02102}} &
\tiny\texttt{F(b1\&F(b2\&F(b3\&Fb4)))\&GF(a1\&X(a2\&X(a3\&Xa4)))} & 13 (\scriptsize 0.74) & 14 (\scriptsize 0.01) & 14 (\scriptsize 0.16) & \textbf{10} (\scriptsize 0.01) \\ &
\tiny\texttt{F(b1\&F(b2\&F(b3\&F(b4\&Fb5))))\&GF(a1\&X(a2\&X(a3\&X(a4\&Xa5))))} & 22 (\scriptsize 0.76) & 23 (\scriptsize 0.04) & 23 (\scriptsize 0.19) & \textbf{14} (\scriptsize 0.03) \\ &
\tiny\texttt{F(b1\&F(b2\&F(b3\&F(b4\&F(b5\&Fb6)))))\&GF(a1\&X(a2\&X(a3\&X(a4\&X(a5\&Xa6)))))} & 39 (\scriptsize 0.77) & 40 (\scriptsize 0.27) & 40 (\scriptsize 0.44) & \textbf{21} (\scriptsize 0.19) \\ &
\tiny\texttt{F(b1\&F(b2\&F(b3\&F(b4\&F(b5\&F(b6\&Fb7))))))\&GF(a1\&X(a2\&X(a3\&X(a4\&X(a5\&X(a6\&Xa7)))\newline )))} & 72 (\scriptsize 0.92) & 73 (\scriptsize 2.99) & 73 (\scriptsize 1.57) & \textbf{34} (\scriptsize 1.76) \\ &
\tiny\texttt{F(b1\&F(b2\&F(b3\&F(b4\&F(b5\&F(b6\&F(b7\&Fb8)))))))\&GF(a1\&X(a2\&X(a3\&X(a4\&X(a5\&X(a6\newline \&X(a7\&Xa8)))))))} & 137 (\scriptsize 1.45) & 138 (\scriptsize 39.11) & 138 (\scriptsize 10.74) & \textbf{59} (\scriptsize 20.84) \\

\bottomrule
\end{tabular}
\caption{Comparison of automata state spaces of \texttt{Owl} and \texttt{Slim} against their reduced versions, \texttt{Owl-Red} and \texttt{Slim-Red}. We report number of states (smallest in bold) and runtime (in seconds). Within \texttt{Owl-Red} and \texttt{Slim-Red} we only measure the time it took to reduce the automaton. We report timeout after 300 seconds. The reduction returns complete automata, with explicit sink state, which is why for some cases, the reduction increases the state size by one.}
\label{tab:appendix-table-1}
\end{sidewaystable}
            
\clearpage

\begin{sidewaystable}[]
\vspace{9.5cm}
\small
\setlength{\tabcolsep}{4pt}
\renewcommand{\arraystretch}{1.1}
\begin{tabular}{
    >{\raggedright\arraybackslash}p{4.5cm}    
    >{\raggedright\arraybackslash}p{10.5cm}    
    >{\centering\arraybackslash}p{1.75cm}     
    >{\centering\arraybackslash}p{1.75cm}     
    >{\centering\arraybackslash}p{1.75cm}     
    >{\centering\arraybackslash}p{1.75cm}     
}
\toprule
  Reference
& Formula
& \texttt{Owl}
& \texttt{Owl-Red}
& \texttt{Slim}
& \texttt{Slim-Red}\\
\midrule

\multirow{4}{*}{\cite{holevcek2004verification}} &
\tiny\texttt{GF((a1\&X!a1)|(!a1\&Xa1)|(a2\&X!a2)|(!a2\&Xa2)|(a3\&X!a3)|(!a3\&Xa3))} & 13 (\scriptsize 0.44) & \textbf{8} (\scriptsize 0.01) & 17 (\scriptsize 0.10) & 10 (\scriptsize 0.01) \\ &
\tiny\texttt{GF((a1\&X!a1)|(!a1\&Xa1)|(a2\&X!a2)|(!a2\&Xa2)|(a3\&X!a3)|(!a3\&Xa3)|(a4\&X!a4)|(!a4\&Xa4)\newline )} & 17 (\scriptsize 0.44) & \textbf{10} (\scriptsize 0.01) & 33 (\scriptsize 0.11) & 18 (\scriptsize 0.13) \\ &
\tiny\texttt{GF((a1\&X!a1)|(!a1\&Xa1)|(a2\&X!a2)|(!a2\&Xa2)|(a3\&X!a3)|(!a3\&Xa3)|(a4\&X!a4)|(!a4\&Xa4)\newline |(a5\&X!a5)|(!a5\&Xa5))} & 21 (\scriptsize 0.48) & \textbf{12} (\scriptsize 0.01) & 65 (\scriptsize 0.20) & 34 (\scriptsize 3.83) \\ &
\tiny\texttt{GF((a1\&X!a1)|(!a1\&Xa1)|(a2\&X!a2)|(!a2\&Xa2)|(a3\&X!a3)|(!a3\&Xa3)|(a4\&X!a4)|(!a4\&Xa4)\newline |(a5\&X!a5)|(!a5\&Xa5)|(a6\&X!a6)|(!a6\&Xa6))} & 25 (\scriptsize 0.62) & \textbf{14} (\scriptsize 0.01) & 129 (\scriptsize 0.69) & 66 (\scriptsize 128.34) \\ &
\tiny\texttt{GF((a1\&X!a1)|(!a1\&Xa1)|(a2\&X!a2)|(!a2\&Xa2)|(a3\&X!a3)|(!a3\&Xa3)|(a4\&X!a4)|(!a4\&Xa4)\newline |(a5\&X!a5)|(!a5\&Xa5)|(a6\&X!a6)|(!a6\&Xa6)|(a7\&X!a7)|(!a7\&Xa7))} & 29 (\scriptsize 1.45) & \textbf{16} (\scriptsize 0.01) & 257 (\scriptsize 3.70) & timeout \\ &
\tiny\texttt{GF((a1\&X!a1)|(!a1\&Xa1)|(a2\&X!a2)|(!a2\&Xa2)|(a3\&X!a3)|(!a3\&Xa3)|(a4\&X!a4)|(!a4\&Xa4)\newline |(a5\&X!a5)|(!a5\&Xa5)|(a6\&X!a6)|(!a6\&Xa6)|(a7\&X!a7)|(!a7\&Xa7)|(a8\&X!a8)|(!a8\&Xa8))} & 33 (\scriptsize 5.93) & \textbf{18} (\scriptsize 0.02) & 513 (\scriptsize 27.97) & timeout \\ &
\tiny\texttt{GF((a1\&X!a1)|(!a1\&Xa1)|(a2\&X!a2)|(!a2\&Xa2)|(a3\&X!a3)|(!a3\&Xa3)|(a4\&X!a4)|(!a4\&Xa4)\newline |(a5\&X!a5)|(!a5\&Xa5)|(a6\&X!a6)|(!a6\&Xa6)|(a7\&X!a7)|(!a7\&Xa7)|(a8\&X!a8)|(!a8\&Xa8)|(\newline a9\&X!a9)|(!a9\&Xa9))} & 37 (\scriptsize 37.75) & \textbf{20} (\scriptsize 0.03) & timeout & -- \\

\midrule

\multirow{12}{*}{\cite{DBLP:journals/jcss/BaierK00023}} &
\tiny\texttt{G("msgsend"->F("acksend"\&(("acKrev")|X(("acKrev")|X(("acKrev")|X("acKrev"))))))} & 9 (\scriptsize 0.49) & \textbf{7} (\scriptsize 0.01) & 37 (\scriptsize 0.10) & 16 (\scriptsize 0.01) \\ &
\tiny\texttt{G("msgsend"->F("acksend"\&(("acKrev")|X(("acKrev")|X(("acKrev")|X(("acKrev")|X("\newline acKrev")))))))} & 19 (\scriptsize 0.50) & \textbf{9} (\scriptsize 0.01) & 77 (\scriptsize 0.12) & 27 (\scriptsize 0.02) \\ &
\tiny\texttt{G("msgsend"->F("acksend"\&(("acKrev")|X(("acKrev")|X(("acKrev")|X(("acKrev")|X(("\newline acKrev")|X("acKrev"))))))))} & 37 (\scriptsize 0.51) & \textbf{15} (\scriptsize 0.01) & 157 (\scriptsize 0.11) & 41 (\scriptsize 0.05) \\ &
\tiny\texttt{G("msgsend"->F("acksend"\&(("acKrev")|X(("acKrev")|X(("acKrev")|X(("acKrev")|X(("\newline acKrev")|X(("acKrev")|X("acKrev")))))))))} & 69 (\scriptsize 0.55) & \textbf{17} (\scriptsize 0.02) & 317 (\scriptsize 0.14) & 65 (\scriptsize 0.17) \\ &
\tiny\texttt{G("msgsend"->F("acksend"\&(("acKrev")|X(("acKrev")|X(("acKrev")|X(("acKrev")|X(("\newline acKrev")|X(("acKrev")|X(("acKrev")|X("acKrev"))))))))))} & 133 (\scriptsize 0.65) & \textbf{19} (\scriptsize 0.04) & 637 (\scriptsize 0.21) & 95 (\scriptsize 0.71) \\ &
\tiny\texttt{G("msgsend"->F("acksend"\&(("acKrev")|X(("acKrev")|X(("acKrev")|X(("acKrev")|X(("\newline acKrev")|X(("acKrev")|X(("acKrev")|X(("acKrev")|X("acKrev")))))))))))} & 261 (\scriptsize 0.86) & \textbf{21} (\scriptsize 0.14) & 1277 (\scriptsize 0.33) & 146 (\scriptsize 3.49) \\ &
\tiny\texttt{G("msgsend"->F("acksend"\&(("acKrev")|X(("acKrev")|X(("acKrev")|X(("acKrev")|X(("\newline acKrev")|X(("acKrev")|X(("acKrev")|X(("acKrev")|X(("acKrev")|X("acKrev"))))))))))))} & 517 (\scriptsize 1.51) & \textbf{24} (\scriptsize 0.60) & 2557 (\scriptsize 0.64) & 202 (\scriptsize 19.72) \\ &
\tiny\texttt{G("msgsend"->F("acksend"\&(("acKrev")|X(("acKrev")|X(("acKrev")|X(("acKrev")|X(("\newline acKrev")|X(("acKrev")|X(("acKrev")|X(("acKrev")|X(("acKrev")|X(("acKrev")|X("acKrev\newline ")))))))))))))} & 1029 (\scriptsize 3.04) & \textbf{25} (\scriptsize 2.25) & 5117 (\scriptsize 1.28) & 283 (\scriptsize 97.48) \\

\midrule

\multirow{7}{*}{NCS} &
\tiny\texttt{GF(a\&X(b)\&X(X(X(c))))} & 8 (\scriptsize 0.46) & 8 (\scriptsize 0.01) & 9 (\scriptsize 0.09) & \textbf{6} (\scriptsize 0.01) \\ &
\tiny\texttt{GF(a\&X(b)\&X(X(X(c)))\&X(X(X(X(d)))))} & 16 (\scriptsize 0.46) & 16 (\scriptsize 0.01) & 17 (\scriptsize 0.09) & \textbf{10} (\scriptsize 0.01) \\ &
\tiny\texttt{GF(a\&X(b)\&X(X(X(c)))\&X(X(X(X(X(d))))))} & 32 (\scriptsize 0.46) & 32 (\scriptsize 0.03) & 33 (\scriptsize 0.09) & \textbf{18} (\scriptsize 0.01) \\ &
\tiny\texttt{GF(a\&X(b)\&X(X(X(c)))\&X(X(X(X(X(X(d)))))))} & 64 (\scriptsize 0.46) & 64 (\scriptsize 0.12) & 65 (\scriptsize 0.10) & \textbf{34} (\scriptsize 0.02) \\ &
\tiny\texttt{GF(a\&X(b)\&X(X(X(c)))\&X(X(X(X(X(X(d))))))\&X(X(X(X(X(X(X(e))))))))} & 128 (\scriptsize 0.52) & 128 (\scriptsize 1.48) & 129 (\scriptsize 0.14) & \textbf{66} (\scriptsize 0.05) \\ &
\tiny\texttt{GF(a\&X(b)\&X(X(X(c)))\&X(X(X(X(X(X(d))))))\&X(X(X(X(X(X(X(X(e)))))))))} & 256 (\scriptsize 0.62) & 256 (\scriptsize 10.96) & 257 (\scriptsize 0.22) & \textbf{130} (\scriptsize 0.15) \\ &
\tiny\texttt{GF(a\&X(b)\&X(X(X(c)))\&X(X(X(X(X(X(d))))))\&X(X(X(X(X(X(X(X(X(e))))))))))} & 512 (\scriptsize 0.71) & 512 (\scriptsize 83.81) & 513 (\scriptsize 0.42) & \textbf{258} (\scriptsize 0.55) \\ &
\tiny\texttt{GF(a\&X(b)\&X(X(X(c)))\&X(X(X(X(X(X(d))))))\&X(X(X(X(X(X(X(X(X(X(e)))))))))))} & 1024 (\scriptsize 1.06) & timeout & 1025 (\scriptsize 0.53) & \textbf{514} (\scriptsize 2.31) \\

\bottomrule
\end{tabular}
\caption{Comparison of automata state spaces of \texttt{Owl} and \texttt{Slim} against their reduced versions, \texttt{Owl-Red} and \texttt{Slim-Red}. We report number of states (smallest in bold) and runtime (in seconds). Within \texttt{Owl-Red} and \texttt{Slim-Red} we only measure the time it took to reduce the automaton. We report timeout after 300 seconds.}
\label{tab:appendix-table-2}
\end{sidewaystable}

\clearpage

\begin{sidewaystable}[]
\vspace{9.5cm}
\small
\setlength{\tabcolsep}{4pt}
\renewcommand{\arraystretch}{1.1}
\begin{tabular}{
    >{\raggedright\arraybackslash}p{4.5cm}    
    >{\raggedright\arraybackslash}p{10.5cm}    
    >{\centering\arraybackslash}p{1.75cm}     
    >{\centering\arraybackslash}p{1.75cm}     
    >{\centering\arraybackslash}p{1.75cm}     
    >{\centering\arraybackslash}p{1.75cm}     
}
\toprule
  Reference
& Formula
& \texttt{Owl-Red}
& \texttt{Slim-Red}
& \texttt{GFM-GF}\\
\midrule
\multirow{4}{*}{\cite{holevcek2004verification}} &
\tiny\texttt{GF((a1\&X!a1)|(!a1\&Xa1)|(a2\&X!a2)|(!a2\&Xa2)|(a3\&X!a3)|(!a3\&Xa3))} & 8 (\scriptsize 0.45) & 10 (\scriptsize 0.11) & \textbf{7} (\scriptsize 0.09) \\ &
\tiny\texttt{GF((a1\&X!a1)|(!a1\&Xa1)|(a2\&X!a2)|(!a2\&Xa2)|(a3\&X!a3)|(!a3\&Xa3)|(a4\&X!a4)|(!a4\&Xa4)\newline )} & 10 (\scriptsize 0.45) & 18 (\scriptsize 0.23) & \textbf{9} (\scriptsize 0.09) \\ &
\tiny\texttt{GF((a1\&X!a1)|(!a1\&Xa1)|(a2\&X!a2)|(!a2\&Xa2)|(a3\&X!a3)|(!a3\&Xa3)|(a4\&X!a4)|(!a4\&Xa4)\newline |(a5\&X!a5)|(!a5\&Xa5))} & 12 (\scriptsize 0.49) & 34 (\scriptsize 4.03) & \textbf{11} (\scriptsize 0.09) \\ &
\tiny\texttt{GF((a1\&X!a1)|(!a1\&Xa1)|(a2\&X!a2)|(!a2\&Xa2)|(a3\&X!a3)|(!a3\&Xa3)|(a4\&X!a4)|(!a4\&Xa4)\newline |(a5\&X!a5)|(!a5\&Xa5)|(a6\&X!a6)|(!a6\&Xa6))} & 14 (\scriptsize 0.63) & 66 (\scriptsize 129.03) & \textbf{13} (\scriptsize 0.09) \\ &
\tiny\texttt{GF((a1\&X!a1)|(!a1\&Xa1)|(a2\&X!a2)|(!a2\&Xa2)|(a3\&X!a3)|(!a3\&Xa3)|(a4\&X!a4)|(!a4\&Xa4)\newline |(a5\&X!a5)|(!a5\&Xa5)|(a6\&X!a6)|(!a6\&Xa6)|(a7\&X!a7)|(!a7\&Xa7))} & 16 (\scriptsize 1.47) & timeout & \textbf{15} (\scriptsize 0.12) \\ &
\tiny\texttt{GF((a1\&X!a1)|(!a1\&Xa1)|(a2\&X!a2)|(!a2\&Xa2)|(a3\&X!a3)|(!a3\&Xa3)|(a4\&X!a4)|(!a4\&Xa4)\newline |(a5\&X!a5)|(!a5\&Xa5)|(a6\&X!a6)|(!a6\&Xa6)|(a7\&X!a7)|(!a7\&Xa7)|(a8\&X!a8)|(!a8\&Xa8))} & 18 (\scriptsize 5.95) & timeout & \textbf{17} (\scriptsize 0.20) \\ &
\tiny\texttt{GF((a1\&X!a1)|(!a1\&Xa1)|(a2\&X!a2)|(!a2\&Xa2)|(a3\&X!a3)|(!a3\&Xa3)|(a4\&X!a4)|(!a4\&Xa4)\newline |(a5\&X!a5)|(!a5\&Xa5)|(a6\&X!a6)|(!a6\&Xa6)|(a7\&X!a7)|(!a7\&Xa7)|(a8\&X!a8)|(!a8\&Xa8)|(\newline a9\&X!a9)|(!a9\&Xa9))} & 20 (\scriptsize 37.78) & timeout & \textbf{19} (\scriptsize 0.23) \\

\midrule

\multirow{2}{*}{\cite{DBLP:conf/atva/Duret-Lutz13}} &
\tiny\texttt{GF((a\&XXXa)|(!a\&XXX!a))} & 18 (\scriptsize 0.44) & 10 (\scriptsize 0.09) & \textbf{7} (\scriptsize 0.08) \\ &
\tiny\texttt{GF((a\&XXXXa)|(!a\&XXXX!a))} & 34 (\scriptsize 0.45) & 18 (\scriptsize 0.10) & \textbf{9} (\scriptsize 0.09) \\ &
\tiny\texttt{GF(a|b)} & 4 (\scriptsize 0.42) & \textbf{1} (\scriptsize 0.09) & \textbf{1} (\scriptsize 0.08) \\

\midrule

\multirow{7}{*}{TDR} &
\tiny\texttt{GF(a\&X(X(X(b))))} & 8 (\scriptsize 0.47) & 6 (\scriptsize 0.11) & \textbf{4} (\scriptsize 0.09) \\ &
\tiny\texttt{GF(a\&X(X(X(X(b)))))} & 16 (\scriptsize 0.47) & 10 (\scriptsize 0.10) & \textbf{5} (\scriptsize 0.08) \\ &
\tiny\texttt{GF(a\&X(X(X(X(X(b))))))} & 32 (\scriptsize 0.48) & 18 (\scriptsize 0.10) & \textbf{6} (\scriptsize 0.09) \\ &
\tiny\texttt{GF(a\&X(X(X(X(X(X(b)))))))} & 64 (\scriptsize 0.49) & 34 (\scriptsize 0.10) & \textbf{7} (\scriptsize 0.08) \\ &
\tiny\texttt{GF(a\&X(X(X(X(X(X(X(b))))))))} & 128 (\scriptsize 0.61) & 66 (\scriptsize 0.14) & \textbf{8} (\scriptsize 0.08) \\ &
\tiny\texttt{GF(a\&X(X(X(X(X(X(X(X(b)))))))))} & 256 (\scriptsize 1.17) & 130 (\scriptsize 0.19) & \textbf{9} (\scriptsize 0.08) \\ &
\tiny\texttt{GF(a\&X(X(X(X(X(X(X(X(X(b))))))))))} & 512 (\scriptsize 4.27) & 258 (\scriptsize 0.48) & \textbf{10} (\scriptsize 0.10) \\ &
\tiny\texttt{GF(a\&X(X(X(X(X(X(X(X(X(X(b)))))))))))} & 1024 (\scriptsize 23.67) & 514 (\scriptsize 0.91) & \textbf{11} (\scriptsize 0.09) \\

\midrule

\multirow{7}{*}{NCS} &
\tiny\texttt{GF(a\&X(b)\&X(X(X(c))))} & 8 (\scriptsize 0.47) & 6 (\scriptsize 0.10) & \textbf{4} (\scriptsize 0.09) \\ &
\tiny\texttt{GF(a\&X(b)\&X(X(X(c)))\&X(X(X(X(d)))))} & 16 (\scriptsize 0.47) & 10 (\scriptsize 0.10) & \textbf{5} (\scriptsize 0.08) \\ &
\tiny\texttt{GF(a\&X(b)\&X(X(X(c)))\&X(X(X(X(X(d))))))} & 32 (\scriptsize 0.49) & 18 (\scriptsize 0.10) & \textbf{6} (\scriptsize 0.08) \\ &
\tiny\texttt{GF(a\&X(b)\&X(X(X(c)))\&X(X(X(X(X(X(d)))))))} & 64 (\scriptsize 0.58) & 34 (\scriptsize 0.11) & \textbf{7} (\scriptsize 0.08) \\ &
\tiny\texttt{GF(a\&X(b)\&X(X(X(c)))\&X(X(X(X(X(X(d))))))\&X(X(X(X(X(X(X(e))))))))} & 128 (\scriptsize 2.00) & 66 (\scriptsize 0.19) & \textbf{8} (\scriptsize 0.09) \\ &
\tiny\texttt{GF(a\&X(b)\&X(X(X(c)))\&X(X(X(X(X(X(d))))))\&X(X(X(X(X(X(X(X(e)))))))))} & 256 (\scriptsize 11.58) & 130 (\scriptsize 0.37) & \textbf{9} (\scriptsize 0.08) \\ &
\tiny\texttt{GF(a\&X(b)\&X(X(X(c)))\&X(X(X(X(X(X(d))))))\&X(X(X(X(X(X(X(X(X(e))))))))))} & 512 (\scriptsize 84.52) & 258 (\scriptsize 0.97) & \textbf{10} (\scriptsize 0.08) \\ &
\tiny\texttt{GF(a\&X(b)\&X(X(X(c)))\&X(X(X(X(X(X(d))))))\&X(X(X(X(X(X(X(X(X(X(e)))))))))))} & timeout & 514 (\scriptsize 2.85) & \textbf{11} (\scriptsize 0.09) \\

\bottomrule
\end{tabular}
\caption{Comparison of automata state spaces of our reduced automata \texttt{Owl-Red} and \texttt{Slim-Red} against our direct construction \texttt{GFM-GF}. We report number of states (smallest in bold) and runtime (in seconds). Within \texttt{Owl-Red} and \texttt{Slim-Red} we report the total time including construction and reduction. We report timeout after 300 seconds.}
\label{tab:appendix-table-3}
\end{sidewaystable}

\end{document}